\documentclass[aps,prb,twocolumn,citeautoscript,superscriptaddress,showpacs,english]{revtex4-1}

\usepackage[version=3]{mhchem}
\usepackage{balance}

\usepackage{graphicx} 
\usepackage{float}
\usepackage[english]{babel}

\usepackage{array}
\usepackage[T1]{fontenc}
\usepackage{amsmath,amssymb}
\usepackage[dvipsnames]{xcolor}
\usepackage{setspace}

\usepackage{comment}
\usepackage{amsmath}
\usepackage{amsfonts}
\usepackage{tabularx}
\usepackage{multirow}
\usepackage{glossaries}

\usepackage[utf8]{inputenc} 
\usepackage{hyperref}

\newacronym{MX}{MX}{macromolecular crystallography}
\newacronym{XRD}{XRD}{X-ray diffraction}
\newacronym{XPS}{XPS}{X-ray photoelectron spectroscopy}
\newacronym{PXRD}{PXRD}{powder X-ray diffraction}
\newacronym{SCXRD}{SCXRD}{single crystal X-ray diffraction}
\newacronym{XFEL}{XFEL}{X-ray free electron laser}
\newacronym{MBA}{MBA}{multibend achromat}
\newacronym{DBA}{DBA}{double-bend achromat}
\newacronym{DTBA}{DTBA}{double triple bend achromat}
\newacronym{TBA}{TBA}{triple-bend achromat}
\newacronym{DFT}{DFT}{density functional theory}
\newacronym{DOS}{DOS}{density of states}
\newacronym{PDOS}{PDOS}{projected density of states}
\newacronym{ROS}{ROS}{reactive oxygen species}
\newacronym{RAP}{RAP}{relative atomic percentage}
\newacronym{APXPS}{APXPS}{ambient X-ray photoelectron spectroscopy}
\newacronym{MFC}{MFC}{mass flow controller}
\newacronym{CL}{CL}{core level}
\newacronym{VB}{VB}{valence band}
\newacronym{VBM}{VBM}{valence band maximum}
\newacronym{ICS}{ICS}{inverse Compton scattering}
\newacronym{SAXS}{SAXS}{small angle X-ray scattering}
\newacronym{PSD}{PSD}{position sensitive detector}
\newacronym{FWHM}{FWHM}{full width half maximum}
\newacronym{UHV}{UHV}{ultra-high vacuum}
\newacronym{DWD}{DWD}{diffraction-weighted dose}
\newacronym{PBE}{PBE}{Perdew-Burke-Ernzerhof}
\newacronym{ORTEP}{ORTEP}{Oak Ridge thermal ellipsoid plot}
\newacronym{CSD}{CSD}{Cambridge structural database}
\newacronym{COD}{COD}{1,5-cyclooctadiene}
\newacronym{RT}{RT}{room temperature}
\newacronym{BE}{BE}{binding energy}
\newacronym{LCAO}{LCAO}{linear combination of atomic orbitals}
\newacronym{TDOS}{TDOS}{total densities of states}
\newacronym{CIF}{CIF}{crystallographic information file}
\newacronym{CCD}{CCD}{charge coupled device}
\newacronym{SMX}{SMX}{small molecule crystallography}
\newacronym{HEWL}{HEWL}{hen egg-white lysozyme}
\newacronym{WCI}{WCI}{wheat chymotrypsin inhibitor}
\newacronym{TcAChE}{TcAChE}{torpedo californica acetylcholinesterase}
\newacronym{GPI}{GPI}{glycinium phosphite}
\newacronym{XAS}{XAS}{X-ray absorption spectroscopy}
\newacronym{TMX}{TMX}{transmission X-ray microscopy}
\newacronym{CT}{CT}{computerised tomography}
\newacronym{XANES}{XANES}{X-ray absorption near-edge structure}
\newacronym{FEOL}{FEOL}{front-end-of-line}
\newacronym{CdTe}{CdTe}{cadmium telluride}
\newacronym{PES}{PES}{photoelectron spectroscopy}
\newacronym{IMFP}{IMFP}{inelastic mean free path}
\newacronym{SCF}{SCF}{self-consistent field}
\newacronym{LDA}{LDA}{local density approximation}
\newacronym{UEG}{UEG}{uniform electron gas}
\newacronym{GGA}{GGA}{generalised gradient approximation}
\newacronym{ASW}{ASW}{augmented spherical waves}
\newacronym{LMTO}{LMTO}{linear muffin-tin orbitals}
\newacronym{LCGO}{LCGO}{linear combination of gaussian orbitals}
\newacronym{PSP}{PSP}{pseudopotential}
\newacronym{USP}{USP}{ultrasoft pseudopotential}
\newacronym{NCP}{NCP}{norm-conserving pseudopotentials}
\newacronym{DLS}{DLS}{Diamond Light Source}
\newacronym{IVU}{IVU}{in-vacuum undulator}
\newacronym{DCM}{DCM}{double-crystal monochromator}
\newacronym{HRM}{HRM}{harmonic rejection mirrors}
\newacronym{MAC}{MAC}{multiple analyser crystals}
\newacronym{TCH}{TCH}{Thompson-Cox-Hastings}
\newacronym{SMXRD}{SMXRD}{small molecule X-ray diffraction}
\newacronym{DE}{DE}{diffraction efficiency}
\newacronym{QE}{QE}{quantum efficiency}
\newacronym{QDE}{QDE}{quantum diffraction efficiency}
\newacronym{NAP}{NAP}{near ambient pressure}
\newacronym{AIMD}{AIMD}{\textit{ab-initio} molecular dynamics}
\newacronym{EF}{EF}{Fermi energy}
\newacronym{ICD}{ICD}{interatomic Coulombic decay}
\newacronym{ETMD}{ETMD}{electron transfer mediated decay}

\begin{document}
    
\title{Simulation of Radiation Damage on \ce{[M(COD)Cl]2} using Density Functional Theory}

\author{Nathalie K.~Fernando}
\affiliation{Department of Chemistry, University College London, 20 Gordon Street, London, WC1H~0AJ, UK.}
\author{Nayera Ahmed}
\affiliation{Department of Chemistry, University College London, 20 Gordon Street, London, WC1H~0AJ, UK.}
\author{Katherine Milton}
\affiliation{London Center for Nanotechnology, University College London, Gower Street, London WC1E 6BT, UK.}
\author{Claire A.~Murray}
\affiliation{Diamond Light Source Ltd, Diamond House, Harwell Science and Innovation Campus, Didcot, Oxfordshire, OX11~0DE, UK.}
\author{Anna Regoutz}
\affiliation{Department of Chemistry, Inorganic Chemistry Laboratory, South Parks Road, OX1~3QR, UK.}
\email{anna.regoutz@chem.ox.ac.uk}
\author{Laura E.~Ratcliff}
\affiliation{Centre for Computational Chemistry, School of Chemistry, University of Bristol, Bristol BS8~1TS, UK.}
\affiliation{Hylleraas Centre for Quantum Molecular Sciences, Department of Chemistry, UiT The Arctic University of Norway, N-9037 Troms\o{}, Norway.}
\email{laura.ratcliff@bristol.ac.uk}

\begin{abstract}

Theoretical calculations of materials have in recent years shown promise in facilitating the analysis of convoluted experimental data. This is particularly invaluable in complex systems or for materials subject to certain environmental conditions, such as those exposed to X-ray radiation during routine characterisation. Despite the clear benefit in this use case to shed further light on intermolecular damage processes, the use of theory to study radiation damage of samples is still not commonplace, with very few studies in existing literature. In this paper, we demonstrate the potential of density functional theory for modelling the electronic structure of two industrially important organometallic systems of the formula \ce{[M(COD)Cl]2} where M=Ir/Rh and COD=1,5-cyclooctadiene, which are subject to X-ray irradiation via X-ray Diffraction and X-ray Photoelectron Spectroscopy. Our approach allows calculated spectra to be compared directly to experimental data, in this case, the X-ray photoelectron valence band spectra, enabling the valuable correlation of individual atomic states to the electronic structure.
\end{abstract}

\maketitle

\section{Introduction}

The simulation of materials, particularly using first principles approaches such as density functional theory (DFT), is invaluable for aiding the analysis of experimental data. This is particularly true where experiment alone may be insufficient to infer information from complex systems under special conditions and phenomena, such as radiation damage. Indeed, the use of theory to study radiation damage is not unheard of. The simulation of radiation-induced structural defects and properties is an active area of research, particularly in nuclear reactor materials,~\cite{Victoria2007, Dudarev2013, Kato2015} DNA and other biologically important systems,~\cite{Sevilla2016} as well as Si particle detectors,~\cite{Passeri2001, Garutti2019, Petasecca2006} which are typically modelled under the irradiation of proton, neutron, gamma, electron, or ion/plasma beams.\par 

Despite this, the application of computational methods is very limited in X-ray irradiation studies. To date, there exist few reports of theoretical models describing X-ray induced sample change. Bernasconi~\textit{et al.}\ studied the X-ray induced effects to a model natural wax crystal, \textit{n}-eicosane, at room temperature, using excited-state \acrfull{AIMD} based on time-dependent density functional theory time-dependent DFT (TD-DFT) at the B3LYP level of theory.~\cite{Bernasconi2016} The authors explored two extreme cases: the first, what is termed the absorption limit, whereby an electron–hole pair is created by promotion of one electron to a low-lying virtual state (possibly the lowest empty state), and the photoemission limit, in which an electron is ejected and a positive hole is created. Close~\textit{et al.}\ used DFT, also with the B3LYP exchange-correlation functional to explain the structural changes induced in crystalline proteins at 100~K. This was achieved by considering specific two one-electron and two-electron reductive and oxidative reaction pathways in a range of amino acids, in gas phase and in solution phase. The aim was to determine which pathway resulted in thermodynamically stable observed products.~\cite{Close2019} Bhattacharyya~\textit{et al.}\ later employed DFT with B3LYP for a select range of protein crystals.~\cite{Bhattacharyya2020} Using theory to complement their single crystal XRD (SCXRD) data, the authors suggested a correlation between the likelihood of the disulfide bond cleavage and the position of a carbonyl O atom along the S—S bond, which provides a pathway for electron transfer for reduction of the bond.\par

These studies clearly showcase the benefit of applying theory to better understand specific reaction pathways and structural damage in organic and macromolecular systems. However, to date, very few studies have used theory to simulate the radiation-induced changes to the electronic structure of a material, not least of a small-molecule crystal. One such example is the work by B\l{}achucki~\textit{et al.}\ who studied the electronic structure changes in \ce{[Fe(CN)^{-6}]4} in solution under X-ray free electron laser (XFEL) irradiation (via non-resonant X-ray emission spectroscopy, NXES) and supported their results with simulations of X-ray emission spectra to explain the occurrence of high oxidation states with increasing flux by removing integer electrons.~\cite{Blachucki2019}  Sch\"urmann~\textit{et al.}\ also combined computational methods with X-ray diffraction studies by applying extended DFT methods in the refinement
of radiation-activated residual density peaks from the BnSeSe radical species.~\cite{Schurmann2022}

Key insights into the chemical states and electronic structure of the transition-metal catalysts upon X-ray irradiation were determined from X-ray photoelectron spectroscopy (XPS), as a complementary technique to XRD in our previous study.~\cite{Fernando2021} In that work, the combined application of XPS to the Ir and Rh complexes, with theoretical calculations based on DFT, was introduced. Good agreement was achieved between the experimental valence band (VB) and the densities of states (DOS) determined by DFT analyses, such that conclusions could be drawn on the contribution of specific states to the electronic structure of the system, in its intrinsic, undamaged state. This enabled hypotheses to be made on the level of contribution of these states after X-ray irradiation by comparison with the VB of the damaged complexes. The benefit of applying theory to these transition-metal catalysts has therefore already been demonstrated.\par 

In this article, we expand on this previous work by using DFT to explore in more detail the electronic structures of \ce{[Ir(COD)Cl]2} and \ce{[Rh(COD)Cl]2} under the effects of X-ray exposure. In doing so we have two main objectives:
\begin{enumerate}
    \item To determine the total and projected densities of states (PDOS) of the complexes after irradiation to gain insights into the effects of irradiation on electronic structure. 
    \item To mimic the physical processes known to occur from XPS under irradiation by applying them to the undamaged structure to see if the resulting spectral changes emulate the experimental VB spectra of the most significantly damaged structure. 
\end{enumerate}
By using the crystal structures of the Ir and Rh complexes after maximum dose irradiation via both XPS and XRD experiments,~\cite{Fernando2021} a close connection between theory and experiment is maintained. In particular, the gradual change in electronic structure observed in the experimental VB with increasing dose is simulated by systematically applying various physical processes known to occur from previous XPS analyses, namely the metal photoreduction and the loss in Cl intensity. The three scenarios of radiation-induced local changes explored are therefore:
\begin{enumerate}
    \item Simulating the photoreduction of the metal
    \item Simulating the loss of Cl$^{-}$
    \item Simulating the concomitant metal reduction and loss of Cl$^{-}$ 
\end{enumerate}

\noindent Studying the effect on the DOS in these extreme (1 and 2) and combined (3) cases enables important similarities to be drawn with the experimental VB changes with increasing X-ray dose. The framework applied here paves the way for future predictive approaches to radiation damage studies of these transition metal complexes.

In this paper, geometry optimisations of the minimum and maximum dose structures for the Ir and Rh complexes offer starting structures for the computational approach outlined in the section below. DOS for the undamaged (minimum dose) and damaged (maximum XRD and XPS dose) structures are compared with the minimum XPS and maximum XPS and XRD dose experimental valence bands, respectively. Following this, we explore a method of explicitly perturbing the system to simulate the effects of radiation and analyse the resulting PDOS.

\section{Methods}

The experimental XPS and XRD methods followed to obtain the minimum and maximum dose structures, XP core-level and valence band spectra, as well as the method of calculating dose using RADDOSE-3D, are outlined in Fernando~\textit{et al.}~\cite{Fernando2021}

\label{sec:2_comp_methods}

\begin{figure*}[!ht]
\includegraphics[width=0.9\textwidth]{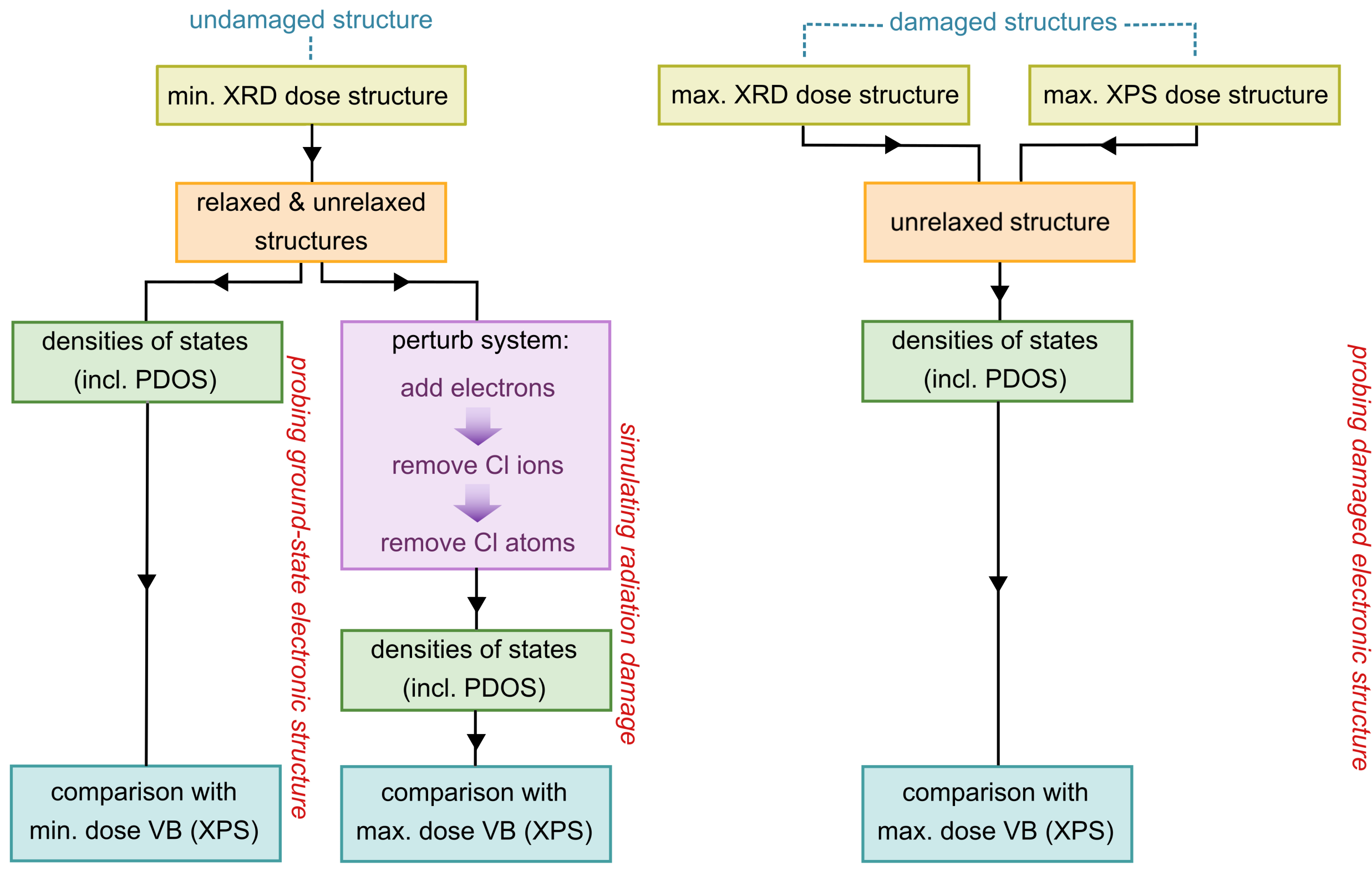}
\caption{\label{fig:exp_flowchart}Schematic summarising the computational approach taken in this study. The starting structures (yellow boxes) input into the calculations were obtained from Rietveld refinements of PXRD patterns collected.} 
\end{figure*}

This section provides an outline of the computational approach, which can be divided into two parts. First, the influence of radiation-induced changes to the atomic structure on the VB spectra were explored by calculating the PDOS of the Ir and Rh complexes for both the undamaged, ground state structures and the experimentally-determined damaged structures. Second, a more detailed exploration of damage was performed by mimicking some of the key physical processes involved. To ensure that the computational results described as closely as possible the true nature of the samples used in experiment, crystal structures were those obtained from Rietveld refinements of XRD patterns of samples. A flowchart summarising the overall theoretical approach is presented in Figure~\ref{fig:exp_flowchart}.\par

\subsection{Influence of Atomic Structure}
\label{subsec:2_b}

Initially, calculations of the total and projected densities of states of \ce{[Ir(COD)Cl]2} and \ce{[Rh(COD)Cl]2} were performed on the ground state structures using the PBE functional~\cite{PBE1996}, and compared directly to experimental VB spectra from XPS experiments. The input crystal structures were obtained from Rietveld refinements of the first minimum XRD dose (corresponding to 6~MGy and 2~MGy for Ir and Rh complexes, respectively), assumed to be the intrinsic, undamaged structure. PDOS calculations were performed for both the unrelaxed (i.e.\ experimental) and relaxed structures, wherein both the atomic positions and cell parameters were optimised. Comparative PDOS calculations were also carried out using the PBE0 hybrid functional.~\cite{Adamo1999} Although PBE0 results were in marginally better agreement with experiment than PBE, the PBE results nonetheless provided satisfactory agreement with experiment and thus PBE is used throughout.\par

Total and projected DOS were then calculated for the damaged structures, using structures obtained from the refinement of PXRD datasets corresponding to an X-ray dose equal to the maximum dose absorbed by the sample during the 35~hour XPS experiment.~\cite{Fernando2021} Similarly, the structures corresponding to the maximum dose achieved at beamline I11 during the PXRD experiments of 2903~MGy and 1012~MGy for the Ir and Rh complexes, respectively, were used to gain insights into the contributions of states at this point of severe structural damage. Only the experimental, unrelaxed structures were employed for this step as the structures reverted back to the undamaged ground state form after geometry optimisation.\par

\subsection{Simulating X-ray Induced Changes}

Following the exploration of structural effects, an attempt to model the effects of radiation damage on the DOS and PDOS of the complexes was made by mimicking the known processes that occur during X-ray exposure, as determined from XPS experiments, namely the photoreduction of the metal and the loss of Cl. These changes were applied to the initial, minimum XRD dose, relaxed and unrelaxed structures, and comparisons made with the maximum XPS dose VB spectra, to determine whether applying these changes theoretically gives a reliable description of the radiation damage process.  In order to simulate the photoreduction of the metal, PDOS calculations were carried out on structures with negative charges explicitly added to the initial structure, from one to 16 in integer electron intervals for the Ir complex, and one to eight for the Rh complex. These values were chosen since they correspond to the total number of metal/chlorine atoms in the associated unit cell.\par 

The next stage was to mimic the severe loss of Cl observed in both Ir and Rh complexes. Since the M—Cl central core comprises of Cl$^{-}$, and in order to remove this ion from the system, Cl atoms from one to 16 were removed in one atom steps from the initial structure, and an extra positive charge added for every Cl atom removed, to take into account the more positively charged complex left behind. As this presents an extreme case, to combine both the metal reduction and the loss of Cl, again, one to 16 and one to eight Cl atoms are removed from the initial Ir and Rh cells, respectively, with no extra charges added to the neutral systems. This combined approach describes the removal of the Cl atom from the system but also the presence of the remaining electron which drives the metal photoreduction at a ratio of 1:1 of Cl atoms lost:electrons added. In order to gain useful insights into the localisation of charge after perturbing the initial (neutral) undamaged system due to adding charges and removing Cl, Bader charges of these new structures were extracted and compared with those of the initial structure.~\cite{Bader1990}

\subsection{Computational Details}

All calculations were performed using the CASTEP code (plane-wave basis set),~\cite{Clark2005} with the PBE exchange-correlation functional with norm-conserving pseudopotentials (\acrshort{NCP}).~\cite{Perdew1996} The number of electrons treated explicitly was 15 and 17 for the metals of \ce{[Ir(COD)Cl]2} and \ce{[Rh(COD)Cl]2}, respectively. A maximum force tolerance of 0.02~eV/\AA~was used for geometry optimisations and the semi-empirical Tkatchenko-Scheffler dispersion correction scheme was applied to account for Van der Waals interactions.~\cite{Tkatchenko2009} \par
All geometry optimisation calculations were carried out with no spin polarisation using the BFGS scheme,~\cite{Pfrommer1997} allowing both lattice parameters and atomic coordinates to be optimised.
For both Ir and Rh complexes, an energy cut-off of 900~eV was selected, whilst a $1\times1\times1$ $\Gamma$--point  only grid was chosen for the Ir complex and a $2\times1\times2$ Monkhorst-Pack \textbf{k}-point grid for the Rh complex.\par
All \acrshort{PDOS} calculations were performed using a Mulliken-style population analysis and later post-processed using the OptaDOS code.~\cite{Morris2014} The density of states were sampled at 0.01~eV intervals. 
For both the Ir and Rh complexes, the LCAO basis states applied were 4/2/2/1 for Ir(Rh)/C/H. 
To account for experimental broadening, a Gaussian broadening of 0.44~eV was applied, determined by the energy resolution of the ThermoScientific K-Alpha spectrometer used in XPS experiments, with a gold foil. \par 
To enable comparison of the DOS with the experimental VB spectra from XPS, the Galore package was used to weight the calculated PDOS,~\cite{Jackson2018} according to the Scofield photoionisation cross-sections at 1.48667~keV (Al K$\alpha$).~\cite{Scofield_1973, Dig_Sco_2020} A Shirley background was subtracted from the experimental VB and normalised relative to the maximum peak intensities. To align the energies of the DOS and valence spectra, the DOS spectra were shifted by a value determined by taking a linear fit of the leading VB edge of the minimum dose VB spectra and determining the energy at half the maximum intensity.

\section{Results \& Discussion}
\label{sec:2_results}
\subsection{Atomic Structure}
\label{subsec:2_e}

Geometry optimisations were carried out on the minimum dose (starting) XRD structures of both Ir and Rh complexes. The lattice parameters obtained from these relaxed structures are presented in Table~S1 in the Supplementary Information along with the lattice parameters from the minimum and maximum dose (start and end) structures,\dag\ extracted from Rietveld refinements of PXRD data, for comparison. The geometry optimised structures clearly show good agreement with the starting experimental structures, with maximum deviations of 2-3\% in the $b$ ($c$) lattice parameter of the Ir (Rh) complex.

\subsection{Ground State Projected Densities of State}
\label{subsec:2_f}

In order to determine the electronic structure of the two complexes at the minimum dose, PDOS calculations were carried out on the relaxed and unrelaxed structures (i.e.\ non-geometry optimised and experimental, respectively). PDOS results obtained directly from these calculations are termed ``unweighted'' as they have not yet been treated to take into account photoionisation cross-sections. \par

Considering the minimum dose structures, at 4~MGy and 2~MGy for Ir and Rh complexes, respectively, there is a region of negative DOS at 5~eV and 13 to 17.5~eV for both complexes, associated primarily with the metal \textit{p}-state. Since negative densities are unphysical, it is likely that the negative DOS is an issue of the Mulliken projection scheme used in the PDOS calculations not being well justified. Earlier tests using the wavelet-based BigDFT code~\cite{Ratcliff2020} to simulate the molecules in gas phase, discussed in Fernando \textit{et al.}\ showed there was no negative PDOS contribution, indicating that this issue is a result of the projection approach, and not a general limitation of DFT or the functional employed to model these systems.~\cite{Fernando2021} 

Another metric for the quality of the projection is the spilling parameter, which gives an indication of whether or not the LCAO basis applied is complete enough.  
Different combinations of LCAO basis functions were tested; in each case the spilling parameter was small ($<$ 1\%), with little variation across basis sets (see Table~S2 in the Supplementary Information).~\dag\ Therefore 4/2/2/1 was chosen in both cases in order to minimise the negative DOS contributions.

\subsubsection{Relaxed vs. Unrelaxed Structures}

Before investigating the changes in electronic structure in the damaged structures, an initial comparison was made of the relaxed \emph{vs.}\ unrelaxed undamaged structures.
To this end, the unweighted PDOS of the relaxed and unrelaxed structures from the minimum dose XRD data are shown in Figures~S1(a) and (b) and Figures~S2(a) and (b), with the corresponding weighted spectra and experimental comparison in Figures~S1(c) and (d) for \ce{[Ir(COD)Cl]2} and Figures~S2(c) and (d) for \ce{[Rh(COD)Cl]2}. The overall lineshapes of the unrelaxed and relaxed structures of the Ir complex have total PDOS that agree well with the experimental VB. The relaxed structure, however, shows the closest agreement, particularly considering the width of the main low energy feature. The unrelaxed PDOS does not take into account the secondary Ir~\textit{d} feature at $\approx$3~eV that is observed both in the relaxed DOS and experimental VB. Previous calculations showed only small differences between the relaxed solid state PDOS structure and that of a gas phase molecule extracted from the structure without further relaxation. Therefore, the differences between the relaxed and unrelaxed spectra can be attributed to differences in internal molecular structure, rather than intermolecular interactions. For example, the Ir-C bond lengths have a much wider spread of values in the unrelaxed structure than in the relaxed one, ranging from 2.09 to 2.11~\AA\ in the relaxed structure compared to 2.05 to 2.29~\AA\ in the unrelaxed structure.  This gives rise to more distinct features in the C $s$ and $p$ peaks in the unweighted PDOS of the unrelaxed structure (see Fig.~S1(c) and (d)), in turn affecting the shape of the weighted PDOS.\par

For the Rh complex, the PDOS of the relaxed structure, shown in Figure~S2(c) has very good agreement with the normalised experimental VB, notably the lineshape of the main Rh~\textit{d} feature at the top of the VB. There are slight discrepancies in the relative intensities between the relaxed theory and experiment in the region between 4--8~eV where there are two competing effects from the Rh~\textit{d} and Cl~\textit{p} states. The differences between the DOS of the unrelaxed structure and the experimental VB, see Figure~S2(d), although subtle, are greater than those between the relaxed structure and the experimental VB. Most notably, there is a slight deviation of the top of the main Rh~\textit{d} state towards lower BE, which results in the misalignment of this region relative to the experiment. This can be attributed to the reduced contribution of the Cl~\textit{p} state at approximately 3~eV in the unrelaxed structure, which leads to an overall underestimation of intensities in this region of the total DOS.\par

Since in both cases the relaxed structures show better agreement with experiment, only these will be used for subsequent comparisons with the results for the damaged structures.

As reported in Fernando~\textit{et al.}, for both complexes, the region closest to the \acrfull{VBM}, is dominated by the metal \textit{d} states, which explains the intensity increase observed in XPS in this region as well as across the 4--8~eV region upon irradiation due to the photoreduction of the metal centres. The PDOS in this energy range shows greater hybridisation with Cl \textit{p} states. From previous XPS results,~\cite{Fernando2021} Cl is known to be lost from both catalysts upon irradiation, however, as there is a notable increase in VB intensity contribution in this region, it can be deduced that the increase in electron density for the metals (i.e.\ photoreduction) dominates.\par

\subsection{PDOS with Damaged Initial Structures}
\label{subsec:2_g}
To confirm the validity of the above hypotheses based on the minimum dose PDOS spectra, and provide further insights into the physical processes upon irradiation, the PDOS of ``damaged'' structures are calculated. These correspond to the structures at maximum absorbed dose during XPS measurements and maximum dose absorbed during the PXRD experiments, which are presented in Table~\ref{tbl:damaged_dose} and obtained from Fernando~\textit{et al}.~\cite{Fernando2021} 

\begin{table}[!h]
\centering
    \caption{X-ray dose of \ce{[Ir(COD)Cl]2} and \ce{[Rh(COD)Cl]2} in their undamaged and damaged states. This corresponds to the minimum XRD dose structure (undamaged) and the maximum dose absorbed at the end of long-duration XPS and PXRD experiments (damaged).} 
    \begin{tabular}{lrr}
\hline \hline
& \multicolumn{2}{c}{Dose / MGy}\\
Experiment condition & \ce{[Ir(COD)Cl]2} & \ce{[Rh(COD)Cl]2}\\
\hline
min. XRD (undamaged) & 4 & 2 \\
max. XPS (damaged) & 26 & 42 \\
max. XRD (damaged) & 2903 & 1012 \\

\hline \hline
    \label{tbl:damaged_dose}
\end{tabular}
\end{table}

\begin{figure*}[!ht]
\includegraphics[width=\textwidth]{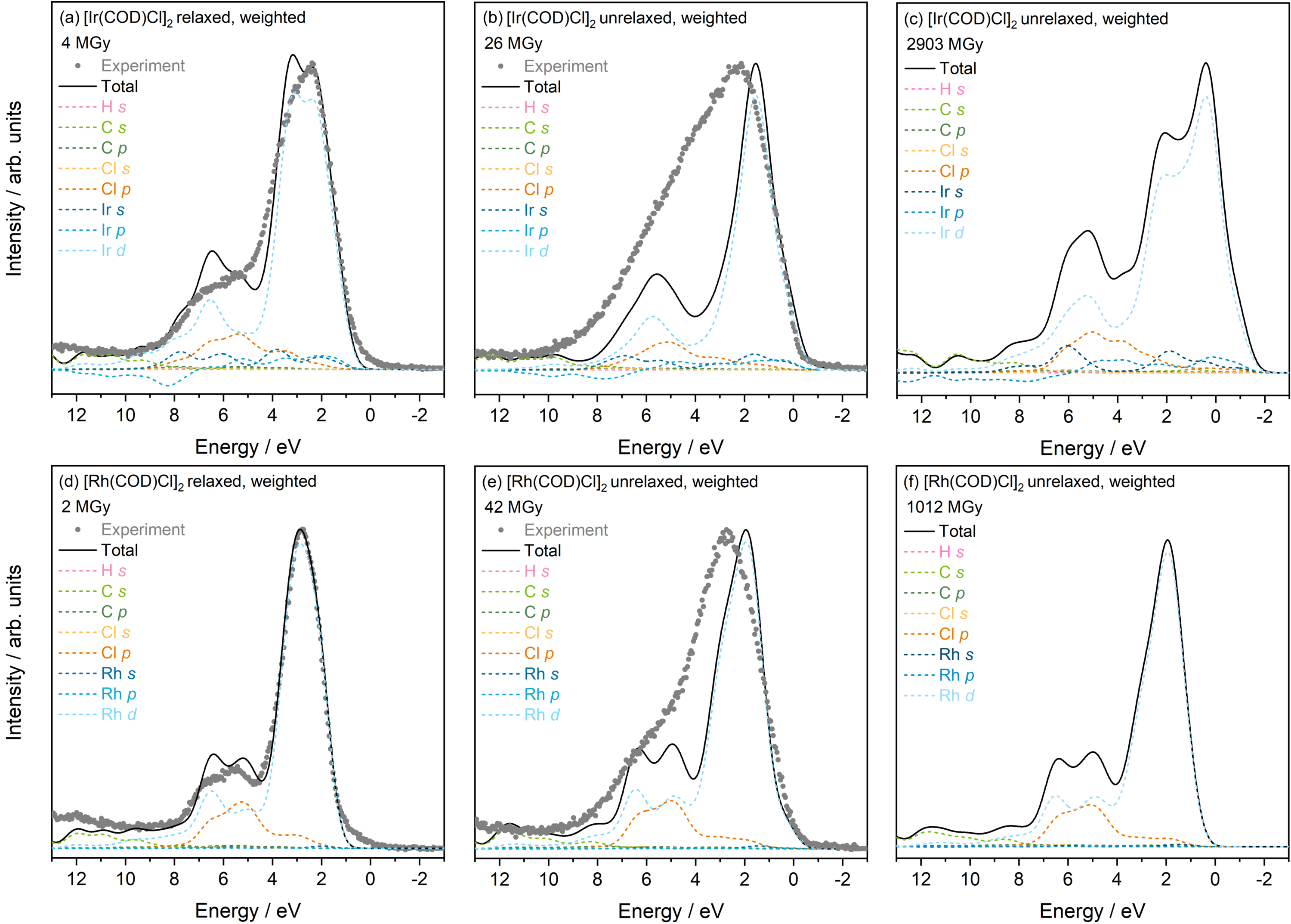}
\caption{\label{fig:Ir_Rh_weighted_pdos_new}Electronic structure of the undamaged (minimum dose) and damaged structures (maximum dose) of \ce{[Ir(COD)Cl]2} and \ce{[Rh(COD)Cl]2}. All subfigures correspond to weighted densities of states of: (a) the minimum XRD dose structure at 4~MGy, (b) at the maximum XPS dose of 26~MGy, and (c) at the maximum XRD dose of 2903~MGy. Similarly, for \ce{[Rh(COD)Cl]2}, the weighted DOS is presented for (d) the minimum XRD dose structure at 2~MGy, (e) at the maximum XPS dose of 42~MGy, and (f) at the maximum XRD dose of 1012~MGy.}
\end{figure*}

PDOS calculations of damaged structures enable direct comparison with the maximum dose VB spectra collected. The unweighted and weighted PDOS for the minimum XRD dose, maximum XPS dose, and maximum XRD dose structures are presented in Figures S3 (in the SI) and Figure~\ref{fig:Ir_Rh_weighted_pdos_new}, respectively. As might be expected due to the lack of external constraints, the damaged Rh structures reverted back to the minimum dose structure, and thus only unrelaxed damaged structures are included.  

Considering first the changes to the PDOS of the Ir complex as X-ray dose is increased, it is clear that there are some discrepancies between the 4~MGy PDOS and TDOS (Figure~\ref{fig:Ir_Rh_weighted_pdos_new}(a)) with that at 26~MGy (Figure~\ref{fig:Ir_Rh_weighted_pdos_new}(b)), the maximum dose achieved during XPS experiments. However, the 4~MGy DOS here is of the relaxed structure, whereas the 26~MGy and 2903~MGy correspond to the PDOS of the unrelaxed structure. In fact, previous work showed that there is no notable difference between the unrelaxed 4~MGy DOS and the unrelaxed 26~MGy DOS~\cite{Fernando2021}. As such, Figure~\ref{fig:Ir_Rh_weighted_pdos_new}(b), shows very little similarity with the VB collected at the same dose. For instance, the overall broadening of the VB, the increase in intensity between 3--6~eV, the drop in intensity of the main Ir~\textit{d} state feature and the slight increase in intensity contributions at the VBM, are not observed in the DOS plot. A further increase in the dose is considered, with the DOS of the system at 2903~MGy, Figure~\ref{fig:Ir_Rh_weighted_pdos_new}(c). It is interesting to note that at this maximum XRD dose, the total DOS shows some resemblance to the experimental VB at maximum XPS dose, despite the disparities in absolute dose. This could indicate that values of absorbed dose during XPS measurements have been underestimated or that the influence of environment plays an important role. At 2903~MGy, there appears to be an increase in contributions at the VBM, as well as at approximately 3~MGy where a secondary feature, associated with the Ir~\textit{d} states, is now prominent. This leads to an overall broadening of the TDOS which is consistent with experimental observations with increasing dose. In the 4--8~eV region, the hybridisation of Ir~\textit{d} and Cl~\textit{p} states remains, although the intensity contributions of both, as well as the Ir~\textit{s} state (at 7~eV) are greater than at 4 and 26~MGy.\par

There are also notable differences in the unweighted and weighted PDOS across the varying dose structures for the Rh complex, see Figures S3 and \ref{fig:Ir_Rh_weighted_pdos_new}(d)-(f). Moving from the 2~MGy PDOS to the maximum XPS dose PDOS at 42~MGy, there is a clear increase in Rh~\textit{d} state intensity contribution at the VBM position, which agrees with the experimental observation, pointing to a metal photoreduction. There appears to be no significant change in the Cl~\textit{p} state at 42~MGy, relative to the minimum dose structure at 2~MGy. However, as in the Ir case, the broadening of the VB at 42~MGy is not completely taken into account in the theory at the same dose. Further increasing the dose to 1012~MGy, does not bring about significant change in the PDOS---the only notable difference being the slight increase in intensity of the Rh~\textit{d} feature at 1.8~eV, which leads to an overall increase in TDOS in this region, as observed in experiment.\par

Overall, direct comparisons of theory and experiment using structures of the damaged Ir and Rh complexes have proved insightful. For both complexes, it is clear that the DOS of the minimum dose, relaxed structure offers the closest agreement to experiment. Increasing the dose to the maximum cumulative dose achieved during XPS experiments, shows some similarity with experiment at the same dose, notably with an increase in DOS towards the VBM, however the extent of these changes appears to be underestimated with respect to experiment. The intensity loss with dose and broadening of the metal \textit{d} state features at this dose is notably less in theory than experiment for both complexes. Although there is no direct experimental VB comparison at the dose achieved at maximum XRD (2903~MGy for Ir and 1012~MGy for Rh), it is evident that for the Ir complex, overall broadening continues, as a result of a growing 3~eV Ir~\textit{d} state feature and increase in DOS at the VBM. From 42~MGy DOS to the maximum dose structure of the Rh complex (1012~MGy), there is surprisingly very little change in DOS, although a slight increase in intensity of the Rh \textit{d} state at 5~eV is observed, which is consistent with the intensity rise observed in experiment. \par

\subsection{Simulating X-ray Induced Chemical Change}
\label{subsec:2_h}
\subsubsection{Simulating X-ray-induced photoreduction}
Given that the use of experimentally damaged structures alone was not sufficient to capture the changes in electronic structure visible in the VB, the next step was to directly model the observed photoreduction by adding a series of negative charges to the minimum dose structures. 

Figures~S4 and S5 in the Supplementary Information show both the unweighted and weighted PDOS of the relaxed and unrelaxed starting minimum dose structures when 1 to 16 and 1 to 8 electrons are added to the Ir and Rh complexes. As discussed above, it is clear that the overall increase in DOS and lineshapes of the relaxed system is closer to the experimental VB than the unrelaxed system. The feature below E\textsubscript{F} at approximately $-$1~eV can be attributed to the unoccupied conduction band states, which are invisible to XPS experiments. The intensity of this peak in the charge neutral case is dependent on how many extra bands are included in the calculation. Increasing the number of electrons in the system, sees a gradual shift of the VBM closer to the E\textsubscript{F}, in addition to a slight intensity loss in the VB (where the main metal Ir~\textit{d} state lies). Both these observations are in good agreement with experiment, however no notable broadening of the TDOS in either the Ir or Rh complexes, with increasing charge is seen in this model, which is reflected in experiment as an increase in intensity contributions between 3.8~eV and 7.0~eV. Based on the fact that the trends observed are consistent in both the relaxed and unrelaxed systems, but the initial DOS spectra agrees closer with the relaxed structures, from now on only the minimum dose relaxed structures of both complexes will be considered.\par

\begin{figure*}[ht]
\centering
\includegraphics[width=1.0\textwidth]{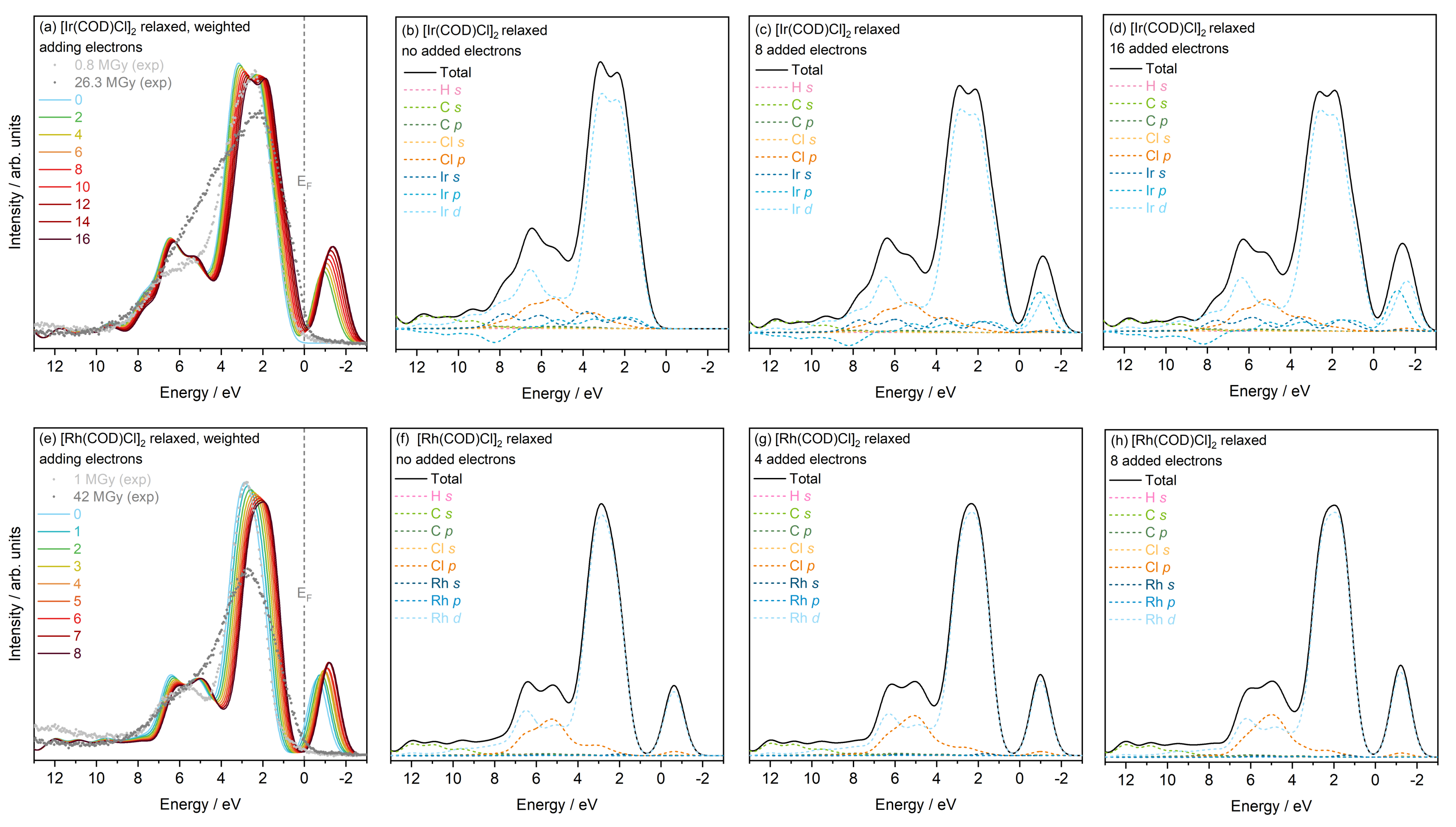}
\caption{\label{fig:Ir_Rh_charges}Calculated density of states for the minimum dose structure of (a)-(d) \ce{[Ir(COD)Cl]2} (top row) and (e)-(h) \ce{[Rh(COD)Cl]2} (bottom row) at 4~MGy and 2~MGy, respectively. (a) the DOS as 1 to 16 charges are added, with the minimum and maximum XPS dose VB. (b)-(d) correspond to the PDOS at various stages of charge addition, with 0 (neutral), 8 and 16 added electrons. (e) the DOS as 1 to 8 charges are added to the Rh complex with the minimum and maximum XPS dose VBs. (f)-(h) the PDOS at various stages of charge addition, with 0 (neutral), 4 and 8 added electrons.}
\end{figure*}

It is interesting to explore the individual contributions of the states to the total DOS when the number of electrons is increased. For simplicity, a comparison is made between no added electrons, 8 (4) added electrons and 16 (8) added electrons for relaxed Ir (Rh) complexes in Figure~\ref{fig:Ir_Rh_charges}. Considering first the top row, corresponding to \ce{[Ir(COD)Cl]2}, there is little to no variation in relative energy positions of all the PDOS features moving from 0 to 8 to 16 added electrons. There is, however, a notable drop in overall intensity of the Ir~\textit{d} state with increasing number of electrons. This is compounded with the subtle drop in the intensity of the first of the split peaks comprising the Ir~\textit{d} contribution, relative to the second, higher energy position feature. Increasing the number of electrons sees no change in the Cl and C state contributions. In the bottom row of Figure~\ref{fig:Ir_Rh_charges} the TDOS and PDOS of the Rh complex, with no added electrons, 4 and 8 added electrons, are presented. As in the Ir calculations, there is a gradual increase in intensity contributions around the VBM and a shift of the VBM towards the \acrfull{EF}. An overall intensity drop is also observed with increasing electrons. From the PDOS Figures~\ref{fig:Ir_Rh_charges}(f)-(h), there is an overall increase in Rh~\textit{d} states towards the VBM, resulting in a more rounded appearance to the main feature of the VB. This change is also observed in experiment. In the 4--8~eV region, where hybridisation of the metal \textit{d} and Cl~\textit{p} states occur, the metal \textit{d}-state intensities gradually drop, which means that the relative contribution of the Cl~\textit{p} state, which has a maximum at approximately 5~eV, increases with increasing electrons. This results in an overall increase in intensity of the TDOS at 5~eV which agrees well with experiment at this energy.\par

\subsubsection{Removing Cl ions}
\begin{figure*}[ht]
\centering
\includegraphics[width=1.0\linewidth]{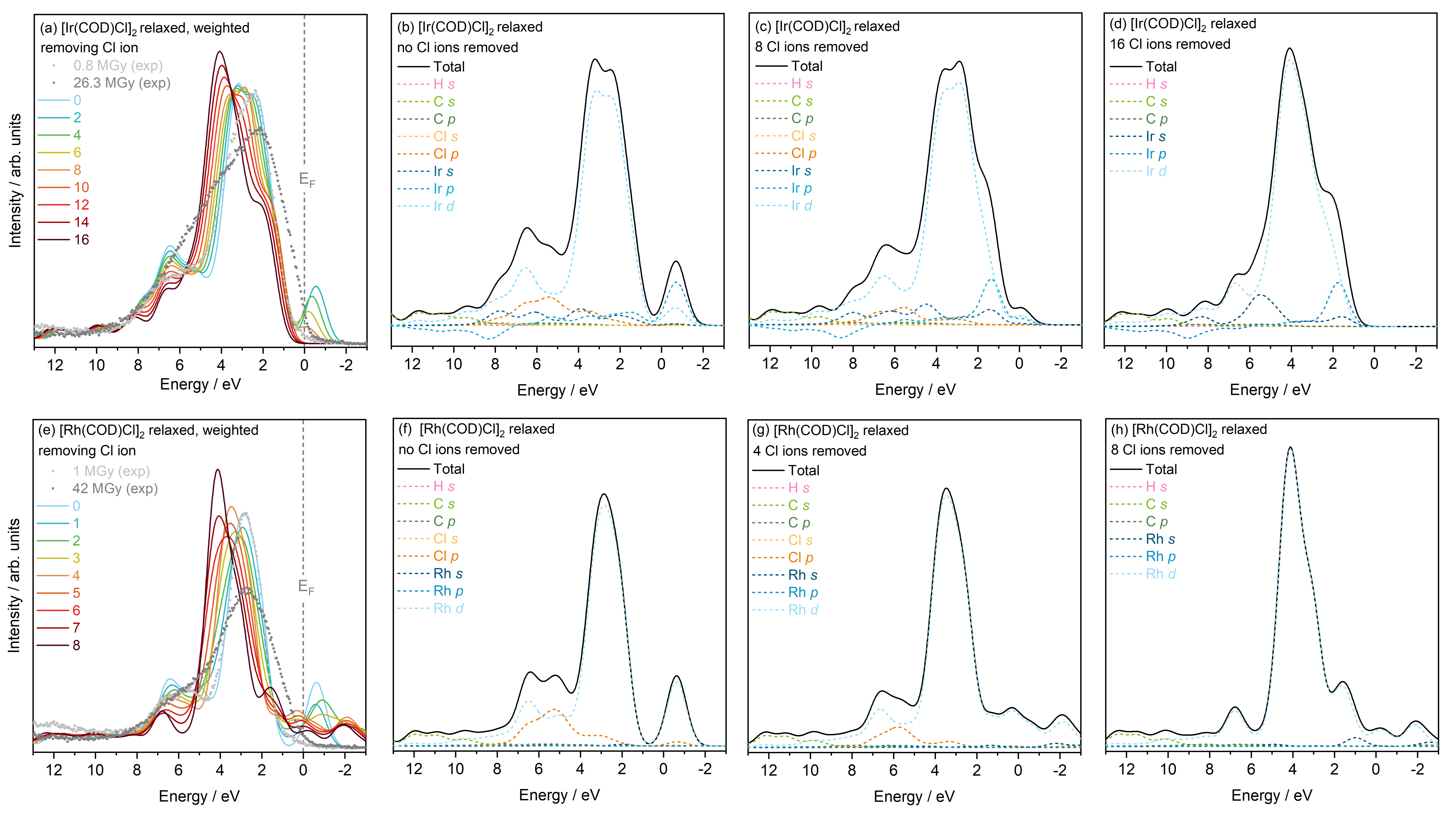}
\caption{\label{fig:Ir_Rh_Cl_ion}Calculated density of states for the minimum dose structure of \ce{[Ir(COD)Cl]2} (top row) and \ce{[Rh(COD)Cl]2} (bottom row) at 4~MGy and 2~MGy, respectively. (a) the DOS as 1 to 16 Cl$^{-}$s are removed, plotted with the minimum and maximum XPS dose VB. (b)-(d) correspond to the PDOS at various stages, after 0, 8 and 16 Cl$^{-}$s removed. (e) the DOS as 1 to 8 Cl$^{-}$s are removed from the Rh complex and plotted with the minimum and maximum XPS dose VBs. (f)-(h) the PDOS at various stages, after 0 (neutral), 4 and 8 removed Cl$^{-}$s.}
\end{figure*}

Aside from the metal photoreduction, another physical process known to occur from the previous radiation XPS studies is the loss of Cl from the complexes. In an attempt to model this process explicitly, 16 and 8 Cl$^{-}$ ions are removed from the minimum dose Ir and Rh complexes, respectively, in integer steps, see Figure~\ref{fig:Ir_Rh_Cl_ion}. First considering the changes in the Ir complex electronic structure (top row), removing the Cl ion \textit{only}, without metal photoreduction, as expected, results in the VBM shifting away from the E\textsubscript{F} towards higher binding energies, contrary to the experimental shift in the VBM towards the E\textsubscript{F}. Interestingly, there is an overall broadening effect on the VB due to increased electron localisation around the Ir \textit{p} and \textit{s} states as more Cl ions are lost.\par 

The electronic structure changes to the Rh complex appear to be more erratic and do not follow a consistent, gradual trend with incremental Cl ion loss. Despite this, the leading edge of the VB (Rh \textit{d} peak) moves away from the E\textsubscript{F} as the number of Cl ions removed reaches 8, which again is contrary to experiment. It becomes increasingly difficult to decipher the point of VBM in Figure~\ref{fig:Ir_Rh_Cl_ion}(e). The intensity contribution of the Rh~\textit{d} is also seen to change erratically. The clear systematic trend that is observable as Cl ions are removed, is the greater localisation of charge around the metal \textit{d} states, reflected in the increasing intensities at approximately 1.5--1.8~eV and at the main feature at 2.5--4.0~eV). In the extreme case of no remaining Cl ions, there appears to be some added charge contribution from the Rh \textit{s} state. Overall, there is very little similarity between experimental VB trends and PDOS when removing Cl ion alone, except the broadening of the Ir complex VB. This model also does not take into account the formation of new Cl species following the cleaving of Cl--M bonds. It is evident that removing Cl$^{-}$ ions alone is too extreme a case. In reality, the competing effects of the metal photoreduction and loss of Cl ions must be considered.\par

\subsubsection{Removing Cl atoms}
\begin{figure*}[ht]
\centering
\includegraphics[width=1.0\linewidth]{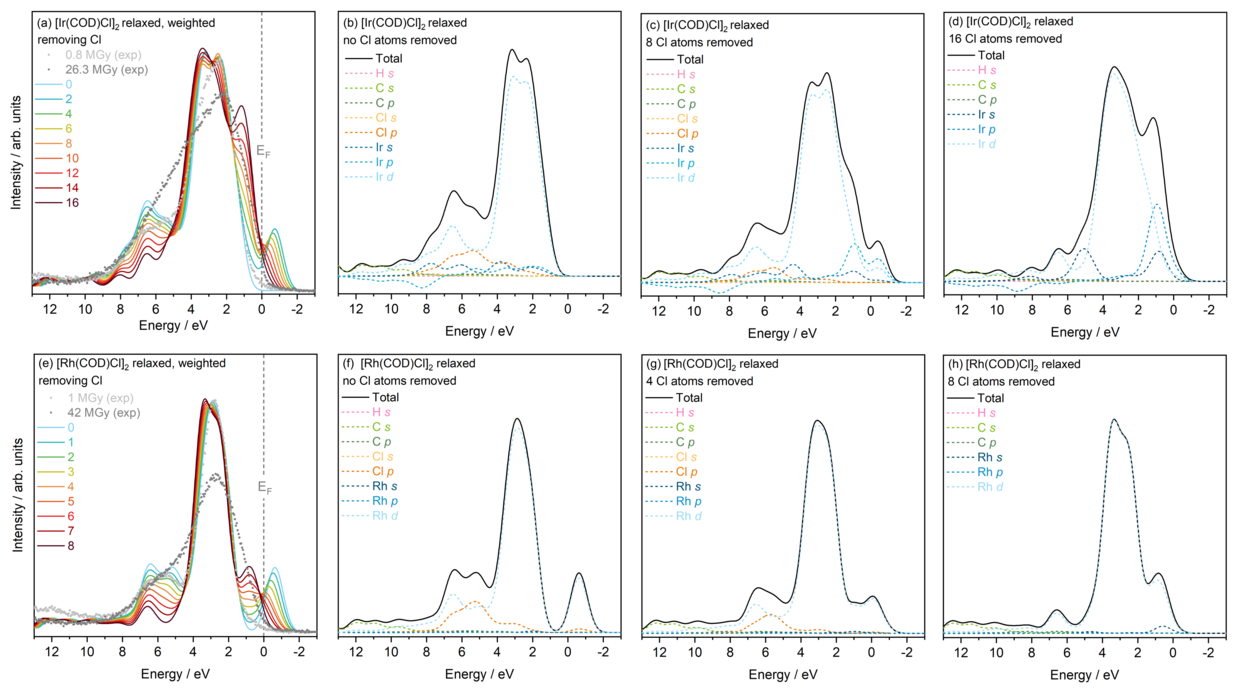}
\caption{\label{fig:Ir_Rh_lossCl}Calculated density of states for the minimum dose structure of (a)-(d) \ce{[Ir(COD)Cl]2} (top row) and (e)-(h) \ce{[Rh(COD)Cl]2} (bottom row) at 4~MGy and 2~MGy, respectively. (a) the DOS as 1 to 16 Cl atoms are removed, plotted with the minimum and maximum XPS dose VB. (b)-(d) correspond to the PDOS at various stages, after 0 (neutral), 8 and 16 Cls removed. (e) the DOS as 1 to 8 Cls are removed from the Rh complex and plotted with the minimum and maximum XPS dose VBs. (f)-(h) the PDOS at various stages, after 0 (neutral), 4 and 8 removed Cls.}
\end{figure*}

Based on the two previous models and to take into account both the metal photoreduction and loss of Cl seen in XPS, Cl atoms as a whole are removed explicitly from the relaxed initial structures, reflecting the loss of the Cl ion, and the excess electron resulting in the reduction of the metal in the M--Cl central core. Again a total of 16 and 8 Cl atoms are removed in integer steps from the Ir and Rh complexes, respectively. The TDOS with increasing Cl loss, along with the PDOS for the neutral case, middle case (half Cl atoms in unit cell removed) and extreme case (all Cl atoms in unit cell removed) are presented in Figure~\ref{fig:Ir_Rh_lossCl}. \par
Overall, there is an increase in DOS towards VBM with Cl atom loss, which can be attributed to the increased localisation of the excess negative charge around the metal \textit{d} states. In the Ir complex, this is also reflected in the broadening of the VB with increasing Cl loss, which is in good agreement with change in the experimental VB with increasing dose. The change observed in the DOS is certainly in closer agreement with the experimental changes, relative to the reduction and Cl-ion loss models discussed previously, as this model takes into account the multi-faceted processes involved. A shifting of Ir \textit{p} states towards higher energies with loss of Cl is also seen to occur, when comparing the neutral, 8 and 16 Cl loss PDOS. This change could point to the Ir \textit{p} state electron localisation and interaction with the Cl atoms i.e.\ the contribution of the Ir \textit{p} state electrons to the bonding with the Cl atoms.\par 
In addition, from 6 to 8~eV, the movement of Ir states depends on whether the electron remains within the system (neutral) or is taken out (Cl ion), affecting the rate of intensity drop in this region. When all the Cl atoms are removed, all the remaining electrons localise around the iridium states which explains the build up of charge near the E\textsubscript{F} in this combined case. This also occurs in the Rh complex as the number of Cl atoms removed increases from 0 to 4 to 8 atoms. Unlike in Ir, only the Rh \textit{d} and Cl \textit{p} states are prevalent. A drop in Cl intensity is observed, as expected, combined with a loss in Rh \textit{d} intensity. As more Cl atoms are removed, charge localisation occurs also on the Rh \textit{s} state. In this particular model, there is a 1:1 loss in Cl to gain in electron. However, from earlier discussions, in the 4--8~eV region, photoreduction dominates, potentially signifying that a 1:2 ratio better describes the scenario. Additionally, it would be interesting to explore relaxation of the structures with added electrons and lost Cl atoms, to determine whether this would match experiment more closely.

\subsubsection{Charge Analysis}
In order to better understand and quantify the localisation of charge after perturbing the minimum dose structure with the addition of electrons and removal of Cl, charge analysis was carried out. Bader and Mulliken charges were first calculated to compare whether both approaches showed similar trends. Figure~S6 shows the change in average Ir and Cl charge as a function of the number of electrons added in the photoreduction model, using both Bader and Mulliken charge analysis approaches. The average change in charge is determined by taking the mean of the charge difference of all Ir/Cl atoms in the unit cell after the addition of the electrons, relative to the neutral case (number of electrons added = 0).\par

It is evident that in both charge analysis models, on average, both the Ir and Cl atoms see a linear increase in charge with electrons added, with the Ir atoms showing the greatest rate of charge increase. The difference in charge is slightly more extreme in the Mulliken charge model with respect to the Bader equivalent, with the Ir atoms consistently showing a significantly greater rate of charge increase and the Cl atoms showing a smaller increase in charge, relative to the Bader model. The differences between Bader and Mulliken charges for the Rh complex are more subtle, see Figure~S7 in the Supplementary Information.\dag\ Given that both Bader and Mulliken show the same trends across both relaxed and unrelaxed initial structures, and as absolute charge values are not particularly important here, rather, the relative comparisons between the M and Cl atoms comprising the central core, only Bader charges will be presented going forward. \par 

\begin{figure*}[!ht]
\centering
\includegraphics[width=0.80\textwidth]{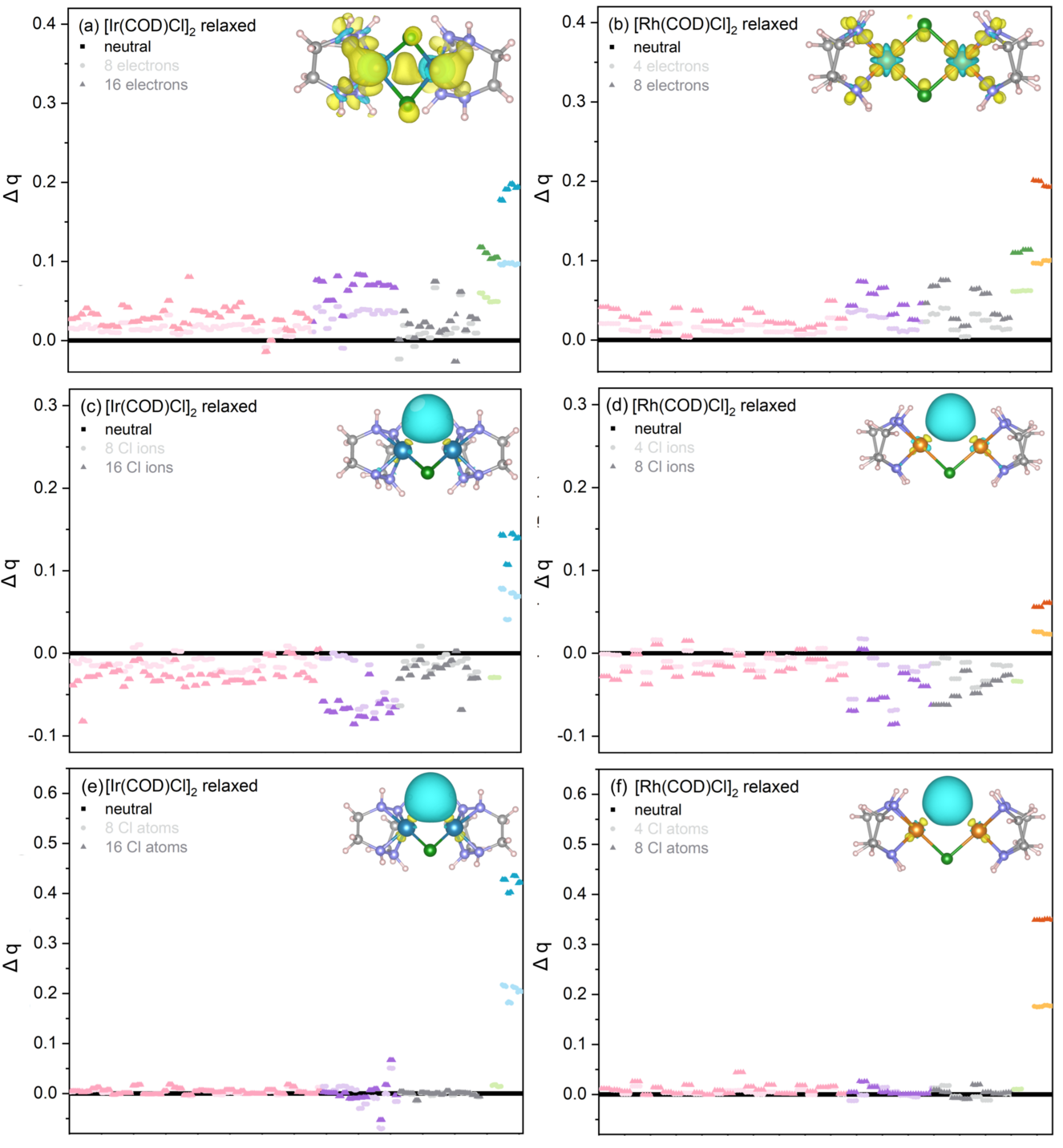}
\caption{\label{fig:Ir_Rh_baders}The change in Bader charge across all atoms in the unit cell, (a) when 8 (light grey) and 16 (dark grey) electrons are added, relative to the neutral (0 added electron) charge for the relaxed, minimum XRD dose structure of \ce{[Ir(COD)Cl]2} and (b) when 4 (light grey) and 8 (dark grey) electrons are added to the \ce{[Rh(COD)Cl]2} structure. (c) and (d) depict the change in Bader charge across all atoms in the unit cell, (c) when 8 (light grey) and 16 (dark grey) Cl ions are removed, relative to the neutral (0 added ions) charge of \ce{[Ir(COD)Cl]2} and (d) when 4 (light grey) and 8 (dark grey) Cl ions are removed from the \ce{[Rh(COD)Cl]2} structure. Similarly, (e) and (f) depict the change in Bader charge across all atoms in the unit cell, (e) when 8 (light grey) and 16 (dark grey) Cl atoms are removed, relative to the neutral (0 removed) charge of \ce{[Ir(COD)Cl]2} and (d) when 4 (light grey) and 8 (dark grey) Cl atoms are removed from the \ce{[Rh(COD)Cl]2} structure. The pink, green and blue/orange data points correspond to the H, Cl, and Ir/Rh atoms, respectively. The purple data points correspond to ligand C, coordinated to the metals and the grey corresponds to C in C--C coordination. The electron density difference, relative to the neutral case, of the individual molecular unit for \ce{[Ir(COD)Cl]2} when 8 electrons/Cl ions/Cl atoms are added/removed, and (d) \ce{[Rh(COD)Cl]2} when 4 electrons/Cl ions/Cl atoms are added/removed are also presented as insets to the corresponding Subfigure. The yellow represents regions of charge addition and cyan, regions of charge depletion.}
\end{figure*}

To gain further insight into the distribution of charge across all atoms in the unit cell, the individual Bader charges are plotted for the neutral, 8 and 16 added electron cases, see Figure~\ref{fig:Ir_Rh_baders}. A clear difference is visible in the charge density difference plots (see inset structures in Subfigures (a) and (b)) with the Rh complex showing more localised changes to the charge density. This is expected from the difference in diffusivity between the Ir~5\textit{d} and Rh~4\textit{d} frontier orbitals, which are more diffuse in the case of Ir, enabling a better overlap with ligand orbitals, thereby leading to a higher degree of covalency. In addition, Ir is less electronegative compared to Rh, again favouring higher covalency. When the number of electrons is increased from 0 to 8 to 16 for the Ir complex (Figure~\ref{fig:Ir_Rh_baders}(a)), the Ir shows the greatest increase in charge, followed by Cl. There is then a notable discrepancy between the charges localised around the C atoms of the cyclooctadiene ligand that are coordinated to other C atoms and the C coordinated to the Ir--Cl centre. Those bonded to the Ir see an overall higher charge density difference, as can be seen in the electron density diagram in the inset of Figure~\ref{fig:Ir_Rh_baders}(a). The Rh complex, see Figure~\ref{fig:Ir_Rh_baders}(b), shows the same order of change in charge from the Rh (highest), Cl, C to H (lowest). However, unlike in the Ir complex, due to the more localised changes to the charge density discussed above which have little impact on either the C coordinated with Rh or other C atoms, there is no notable difference in the charge between the two different coordinations of C. In both complexes, doubling the number of added electrons results in the doubling of absolute charge localised around all atoms in the unit cell.\par

Increasing the number of Cl ions removed from 0 to 8 to 16 for the Ir complex (Figure~\ref{fig:Ir_Rh_baders}(c)) and 0 to 4 to 8, for the Rh complex (Figure~\ref{fig:Ir_Rh_baders}(d)), shows the metals to have the greatest increase in charge. In fact, as the negative charge of Cl$^{-}$ is removed, all other atoms (H and C) experience a notable drop in negative charge, although this drop is small, relative to the loss of electron density from the Cl and small gain around the metals, as is highlighted in the schematics of the electron densities of the molecular unit in the inset of Figures \ref{fig:Ir_Rh_baders}(c) and (d). In other words, the system overall becomes more positively charged, as is expected.\par

In the combined photoreduction and Cl loss case, increasing the number of Cl atoms removed from 0 to 8 to 16 for the Ir complex (Figure~\ref{fig:Ir_Rh_baders}(e)) and 0 to 4 to 8, for the Rh complex (Figure~\ref{fig:Ir_Rh_baders}(f)), again shows the metals to have a substantial increase in negative charge. This can be attributed to the remaining negative charge of Cl$^{-}$. These results show that this excess charge from the Cl localises predominantly around the metals after M--Cl bond cleavage. All other atoms (H and C) comprising the molecule experience almost no change in charge, relative to the neutral case as Cl atoms are removed. These results are clearly presented in the electron density maps, see insets of Figures~\ref{fig:Ir_Rh_baders}(e) and (f).\par

\begin{figure*}[!ht]
\centering
\includegraphics[width=\textwidth]{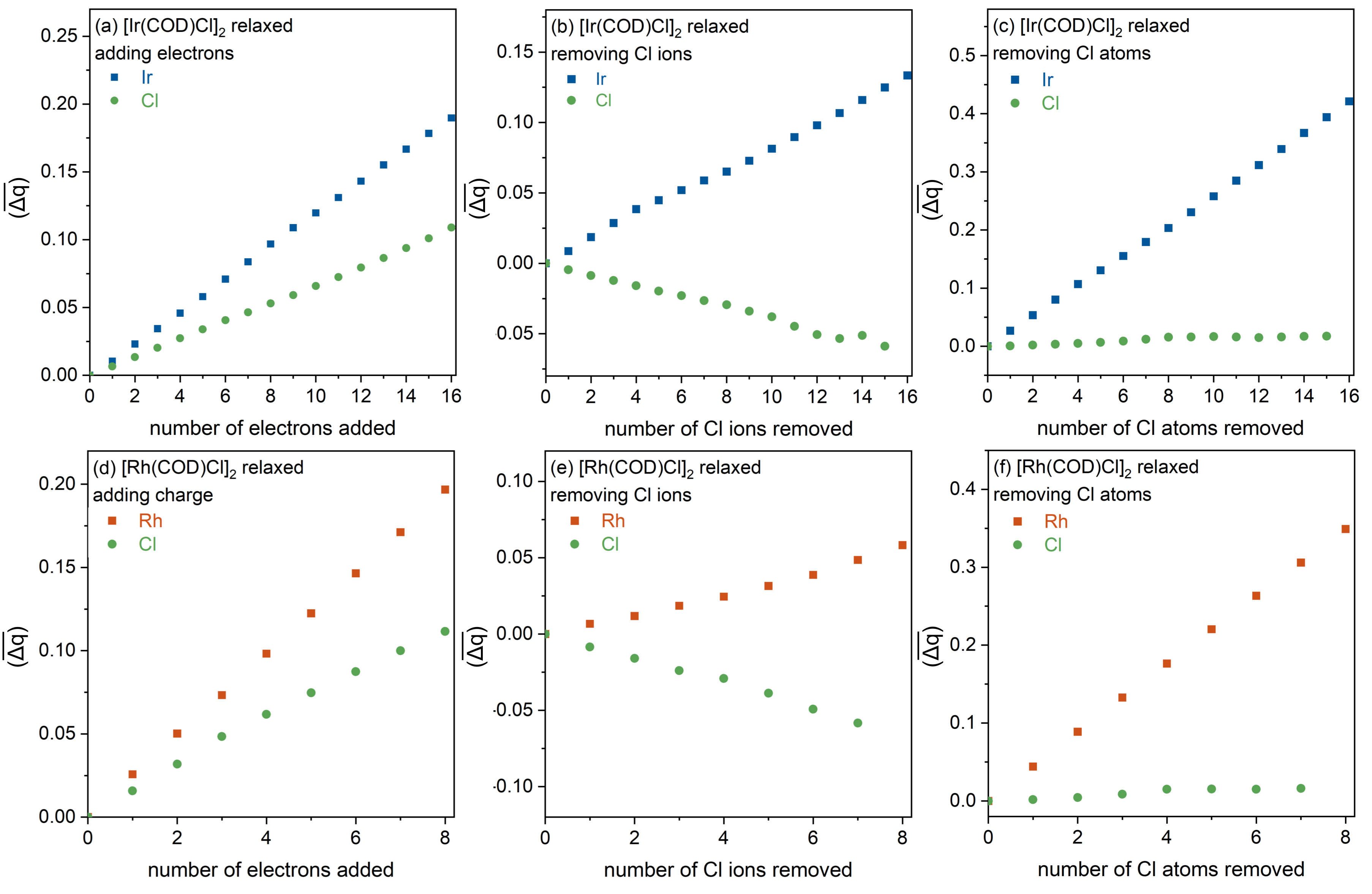}
\caption{\label{fig:rel_bader}The change in Bader charge averaged across all metal and Cl atoms in the unit cell relative to the neutral charge case for the relaxed, minimum XRD dose structures of the two complexes. The top row corresponds to \ce{[Ir(COD)Cl]2} when (a) electrons are added, (b) Cl ions are removed, and (c) the combined case when Cl atoms are removed. The bottom row, (d)-(f), are the equivalent plots for \ce{[Rh(COD)Cl]2}.}
\end{figure*}

The average change in Bader charge is again extracted, as in Figure~S6, but now explored also as a function of electrons added, and Cl ions and Cl atoms removed, focusing only on the M and Cl atom charge contributions.
In the photoreduction scenario alone, for both complexes, the metal and Cl atoms see a gradual increase in average charge with increasing electrons, see Figures~\ref{fig:rel_bader}(a) and (d) for Ir and Rh complexes, respectively. This observation confirms that the electrons localise predominantly around the metals (which experience a steeper incline in charge with added electrons) but also around the Cl atoms. Next, when removing the Cl ions alone, see Figures~\ref{fig:rel_bader}(b) and (e), the metals experience a gradual linear increase in charge, accompanied by a simultaneous loss in overall charge density around the Cl atoms, which can be attributed to the remaining excess negative charge localising around the metals as would be expected from the cleavage of the Cl--M bond. The rate of this charge increase, however, is small, as the system experiences an overall positive charge. For the Rh complex the loss of Cl charge is approximately equal to the gain in Rh charge with Cl ion loss, although again, these changes are very subtle. Removing the neutral Cl atom, see Figures~\ref{fig:rel_bader}(c) and (f), the localisation of charge around the Cl atoms remains almost constant with increasing Cl atom loss. Meanwhile, both Ir and Rh metals see a large increase in charge, which can be attributed to the additional remaining electron from the Cl$^{-}$.

\section{Conclusions}
\label{sec:2_concl}

In recent years, the value of predicting experimental XP spectra using computational calculations has gained recognition, particularly in the calculation of core-level spectra.~\cite{Vines2018, Pi2020, Regoutz2020, Crotti2007} In this work, the combined use of theory and experiment was extended to determine X-ray-induced electronic structure changes using DFT and XPS to aid in the deconvolution of experimental valence band spectra. In so doing, the potential to uncover insights into chemical changes (photoreduction and loss of ligands) to complex organometallic systems with increasing X-ray dose was demonstrated. 

Specifically, a direct comparison of calculated spectra to experimental X-ray photoelectron valence band spectra enabled the correlation of individual atomic states to the electronic structure via the PDOS. These calculations allowed for an enhanced understanding of the electronic structure of damaged structures from the end of previous long-duration XPS and XRD measurements, acquired from refinement of PXRD data at the respective absorbed dose. In addition, the use of theory to mimic physical processes known to occur from XPS studies in incremental steps shed further light on the influence of photoreduction and loss of Cl on the electronic structure of the catalysts. The impact on the electron density via complementary Bader charge analysis successfully enabled the quantification of atomic charge localisation at various stages of the damage process. The work presented here confirms that adding charge and removing Cl from the Ir and Rh complexes alone is very much an extreme case, and that in order to more accurately simulate experimental spectra with DFT, a combination of these competing effects must be taken into account. \par

The scenarios considered here are clearly not exhaustive of the radiation-induced processes that occur. However, these systematic calculations pave the way for future studies incorporating other physical effects. For instance, future calculations could account for the photoelectron-induced presence/formation of cascading radicals, in addition to the formation of new Cl species, known to occur from previous XPS experiments reported in Fernando~\textit{et al.}~\cite{Fernando2021} Additionally, to better understand effects of ambient sample environments on X-ray damage, systematic DOS calculations of the system under ambient conditions (alongside comparative ambient pressure XPS experiments) could be carried out. Finally, the combination of theory and experiment employed here could be further extended to other small molecule crystal systems, where a detailed exploration of electronic structure can provide invaluable information to better understand their intrinsic behaviour under X-ray irradiation.

\section*{Data availability}

Data for this article, including CIF information for key experimentally determined structures, PDOS, XPS data, and charge analysis results for all figures in the main manuscript, are available at Zenodo in Origin format at https://doi.org/10.5281/zenodo.16985474.

\section*{Conflicts of interest}
There are no conflicts to declare.

\section*{Acknowledgements}
NKF acknowledges support from the Engineering and Physical Sciences Research Council (EP/L015277/1). LER acknowledges support from an EPSRC Early Career Research Fellowship (EP/P033253/1).  This work was carried out with the support of Diamond Light Source, instrument I11 (proposal EE19420). The authors acknowledge the use of the UCL Kathleen High Performance Computing Facility (Kathleen@UCL), the UCL Myriad High Performance Computing Facility (Myriad@UCL), and associated support services, in the completion of this work.





\bibliography{ref} 
\bibliographystyle{rsc} 

\end{document}


\centering\noindent \Large{\textbf{{Simulation of Radiation Damage on \ce{[M(COD)Cl]2} using Density Functional Theory}\\
Supplementary Information}}\\

\vspace{7.5mm}

\noindent\large{Nathalie K.~Fernando,$^{\ast}$\textit{$^{a}$}, Nayera Ahmed,\textit{$^{a}$} Katherine Milton,\textit{$^{b}$} Claire A.~Murray,\textit{$^{c}$} Anna Regoutz\textit{$^{a,d}$}} and Laura E.~Ratcliff,\textit{$^{e,f}$} \\
\vspace{0.5cm}

\noindent\small
\textit{$^{a}$~Department of Chemistry, University College London, 20 Gordon Street, London, WC1H~0AJ, UK.}\\
\textit{$^{b}$~London Center for Nanotechnology, University College London, Gower Street, London WC1E 6BT, UK.}\\
\textit{$^{c}$~Diamond Light Source Ltd, Diamond House, Harwell Science and Innovation Campus, Didcot, Oxfordshire, OX11~0DE, UK.}\\
\textit{$^{d}$~Department of Chemistry, Inorganic Chemistry Laboratory, South Parks Road,OX1~3QR, UK. anna.regoutz@chem.ox.ac.uk}\\
\textit{$^{e}$~Centre for Computational Chemistry, School of Chemistry, University of Bristol, Bristol BS8~1TS, UK.}\\
\textit{$^{f}$~Hylleraas Centre for Quantum Molecular Sciences, Department of Chemistry, UiT The Arctic University of Norway, N-9037 Troms\o{}}.\\

\vspace{1cm}

\captionsetup[table]{labelfont=bf,textfont={stretch=1.125,small,sf},labelfont=bf,labelsep=space}

\begin{table*}[!h]
\caption{Unit cell parameters obtained from the Rietveld refinement of the first (PXRD\textsubscript{start}) and final (PXRD\textsubscript{end}) PXRD data and from DFT calculations for \ce{[Ir(COD)Cl]2} and \ce{[Rh(COD)Cl]2}. $a$, $b$, and $c$ are the lattice parameters, $\alpha$, $\beta$, and $\gamma$ are the unit cell angles, and $V$ is the unit cell volume. The percentage difference ($\Delta$\%) between the first experimental and theoretical values is also included. It should noted that the theoretical lattice parameters are marginally different to those first quoted in Fernando~\textit{et al.} (less than 2\% and 0.5\% difference in the crystal volumes for the Iridium and Rhodium complexes, respectively) due to minor changes in the simulation setup.~\cite{Fernando2021}}
\label{tab:exp_vs_theory2}
\begin{tabularx}{\linewidth}{p{1.7cm}p{1.5cm}p{1.5cm}p{1.5cm}p{1.5cm}p{1.5cm}p{1.5cm}p{1.5cm}}\hline\hline
& $a$ / \AA & $b$ / \AA & $c$ / \AA & $\alpha$ / $^{\circ}$ & $\beta$ / $^{\circ}$ & $\gamma$ / $^{\circ}$ & $V$ / \AA$^{3}$\\ \hline
Ir(COD)Cl{]}\textsubscript{2}\\
DFT & 15.216 & 13.278 & 16.0327 & 90.000 & 90.000 & 89.998 & 3239.17\\
PXRD\textsubscript{start} & 15.1775(1) & 13.6407(1) & 16.3046(2)  & 90.000 & 90.000 & 90.000 & 3375.56\\
PXRD\textsubscript{end} & 15.2883(5) & 13.7562(5) & 16.3408(6) & 90.000 & 90.000 & 90.000 & 3436.62 \\
$\Delta$\% & -0.254 & -2.660 & -1.668 & 0.000 & 0.000 & -0.002 & -4.04\\

\hline
Rh(COD)Cl{]}\textsubscript{2}\\
DFT & 7.210 & 25.473 & 8.772 & 90.000 & 91.101 & 90.000 & 1610.823\\
PXRD\textsubscript{start} & 7.3008(1) & 25.4525(4) & 9.0483(1)  & 90.000 & 91.877(1) & 90.000 & 1680.46\\
PXRD\textsubscript{end} & 7.3011(1) & 25.4869(5) & 9.0642(2) & 90.000 & 91.988(1) & 90.000 & 1685.67\\
$\Delta$\% & -1.244 & 0.080 & -2.996 & 0.000 & -0.844 & 0.000 & -4.14\\
 \hline\hline
 \label{tab:exptheo2}
\end{tabularx}
\end{table*}

\begin{table}[!h]
\centering
    \caption{The spilling parameters output from the calculations of \ce{[Ir(COD)Cl]2} and \ce{[Rh(COD)Cl]2}, when different linear combination of atomic orbital (LCAO) parameters are used.} 
    \begin{tabular}{crr}
\hline \hline
LCAO states  &  \multicolumn{2}{c}{spilling parameter / \%}  \\
  & \ce{[Ir(COD)Cl]2} & \ce{[Rh(COD)Cl]2} \\
\hline
default & 0.83 & 0.72 \\
4221 & 0.83 &  0.84 \\
4321 & 0.78 &  0.84 \\
5221 & 0.77  &  0.72 \\
5321 & 0.73  & - \\
6221 & 0.57  & 0.67 \\
\hline \hline
    \label{tbl:spillage_param}
\end{tabular}
\end{table}

\begin{figure*}[!h]
\centering
\includegraphics[width=0.8\textwidth]{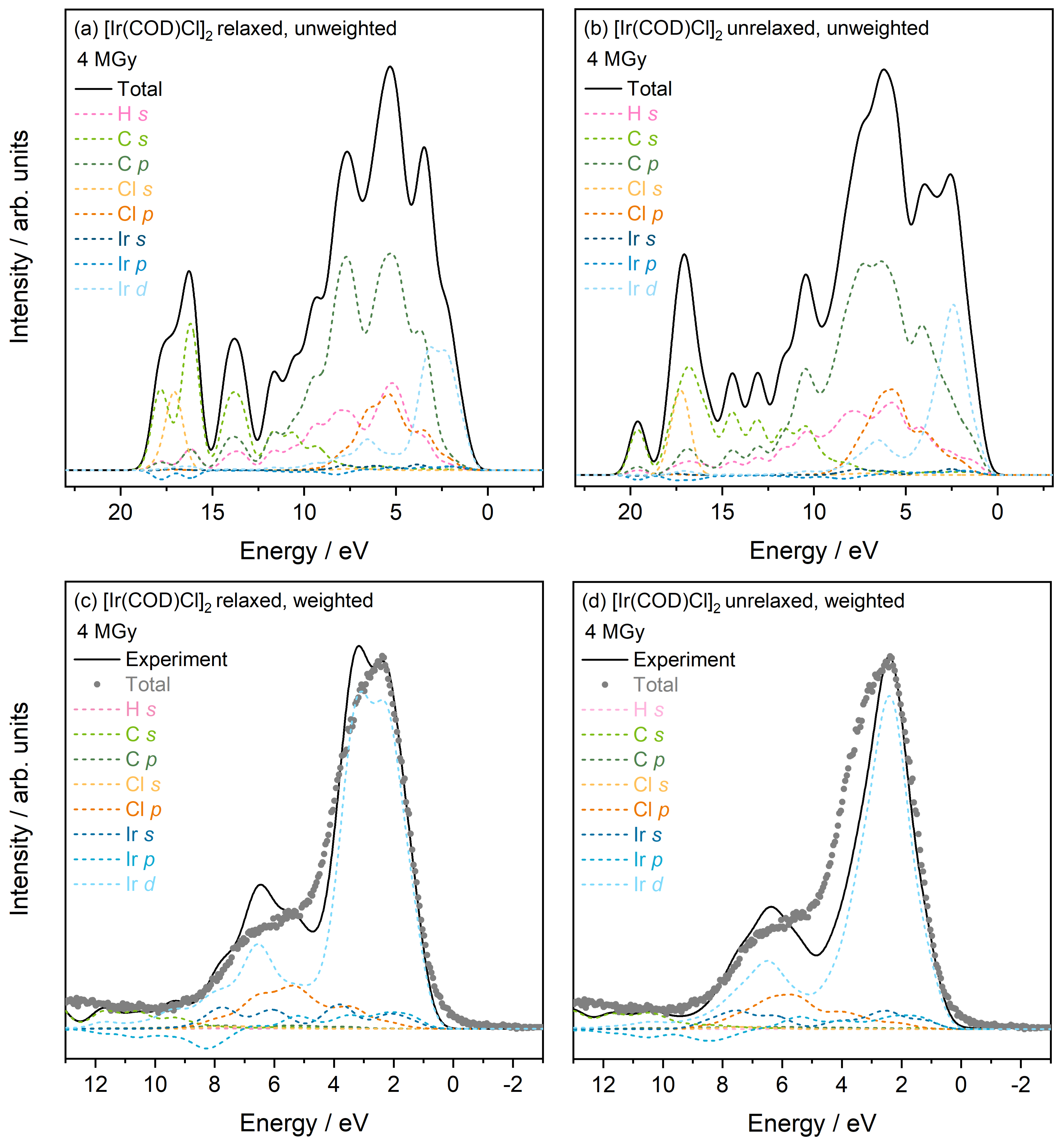}
\caption{\label{fig:Ir_un_relaxed}Electronic structure of the \ce{[Ir(COD)Cl]2} catalyst. (a) and (b) the unweighted DOS for the relaxed and unrelaxed system, ground state systems, respectively. (c) and (d) the broadened and photoionisation cross section corrected total (TDOS) and projected (PDOS) density of states, calculated using the CASTEP code with the PBE functional, compared to the initial minimum dose XPS VB data.}
\end{figure*}

\begin{figure*}[!ht]
\centering
\includegraphics[width=0.8\textwidth]{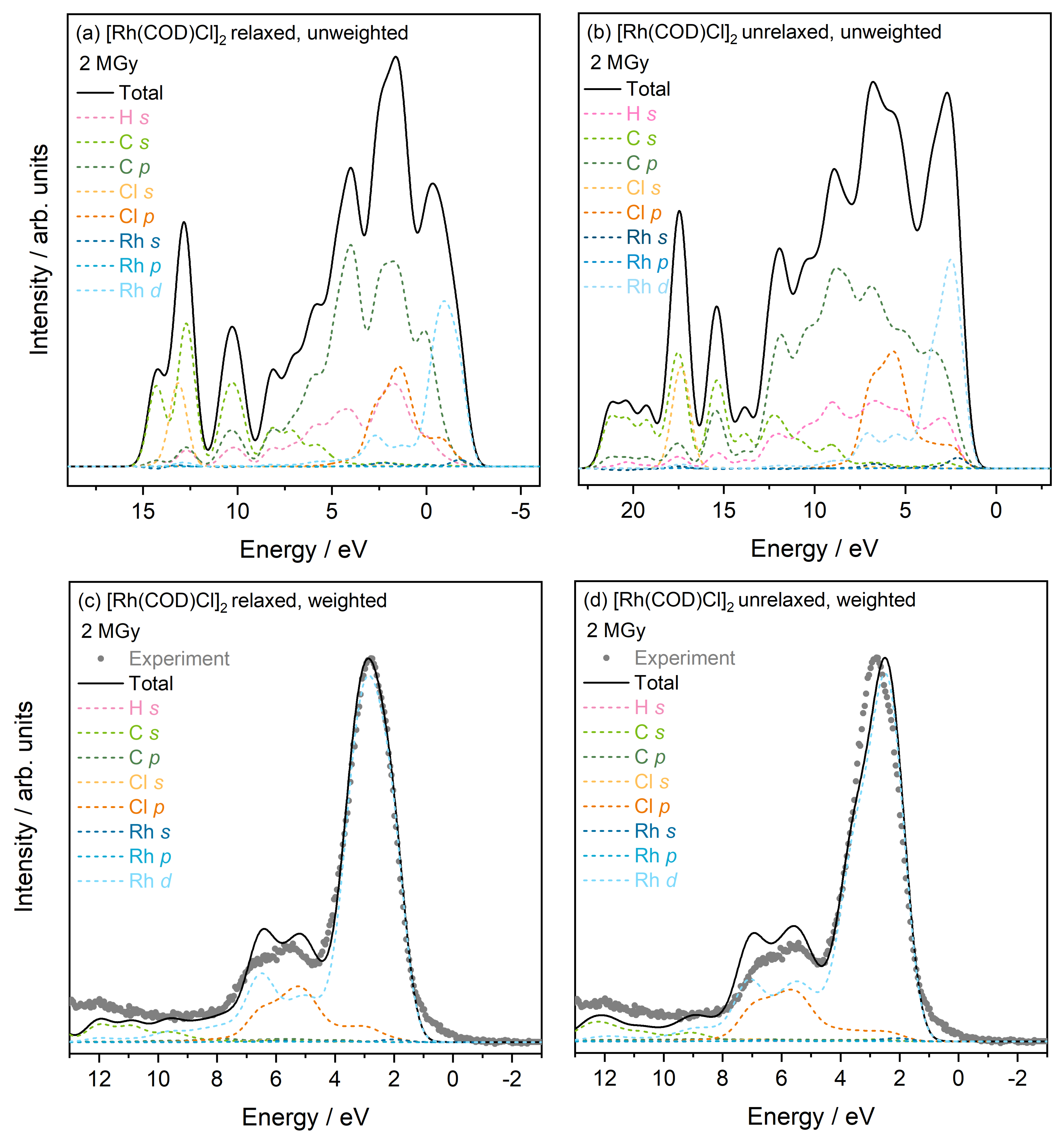}
\caption{\label{fig:Rh_un_relaxed}Electronic structure of the \ce{[Rh(COD)Cl]2} catalyst. (a) and (b) the unweighted DOS for the relaxed and unrelaxed system, ground state systems, respectively. (c) and (d) the broadened and photoionisation cross section corrected total (TDOS) and projected (PDOS) density of states, calculated using the CASTEP code with the PBE functional, compared to the initial minimum dose XPS VB data.}
\end{figure*}

\begin{figure*}[!h]
\includegraphics[width=\textwidth]{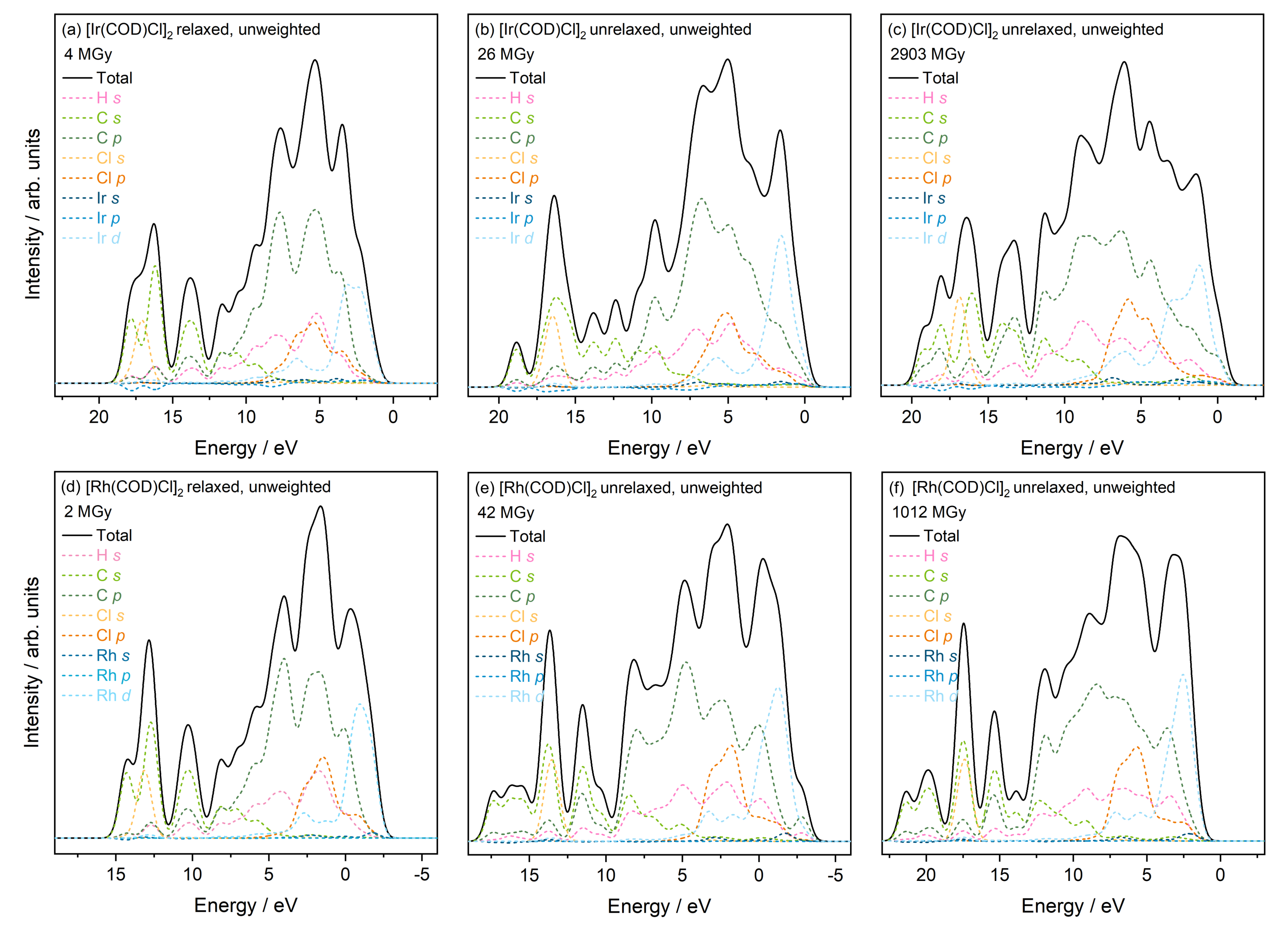}
\caption{\label{fig:Ir_Rh_unweighted_pdos_new}Electronic structure of the undamaged (minimum dose) and damaged structures (maximum dose) of \ce{[Ir(COD)Cl]2} (top row) and \ce{[Rh(COD)Cl]2} (bottom row). All subfigures correspond to unweighted densities of states of: (a) the minimum XRD dose structure at 4~MGy, (b) for the structure at maximum XPS dose of 26~MGy, and (c) for the structure at the maximum XRD dose of 2903~MGy. Similarly, for \ce{[Rh(COD)Cl]2}, the unweighted DOS is presented for (d) the minimum XRD dose structure at 2~MGy, (e) at the maximum XPS dose of 42~MGy, and (f) at the maximum XRD dose of 1012~MGy.}
\end{figure*}

\begin{figure*}[!h]
\centering
\includegraphics[width=0.75\textwidth]{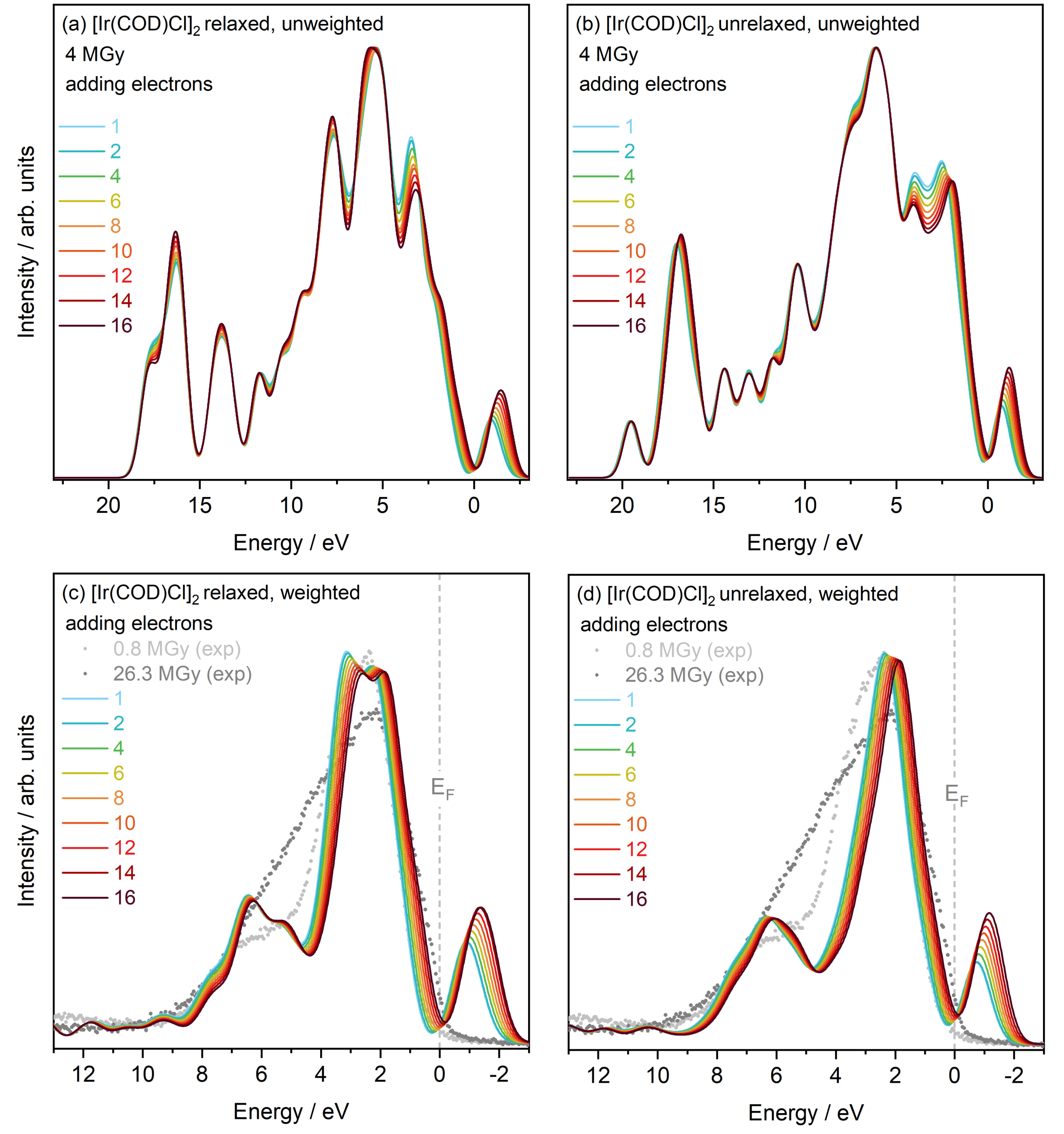}
\caption{\label{fig:Ir_reduction}The calculated density of states for the minimum XRD dose structure of \ce{[Ir(COD)Cl]2} at 4~MGy, mimicking the photoreduction process as 1 to 16 electrons are added to the system. (a) and (b) correspond to the unweighted DOS, for the relaxed and unrelaxed initial structures, respectively. (c) and (d) are the weighted DOS of the relaxed and unrelaxed Ir complex, plotted alongside the minimum and maximum XPS dose VB spectra for comparison.}  
\end{figure*}

\begin{figure*}[!hb]
\centering
\includegraphics[width=0.75\textwidth]{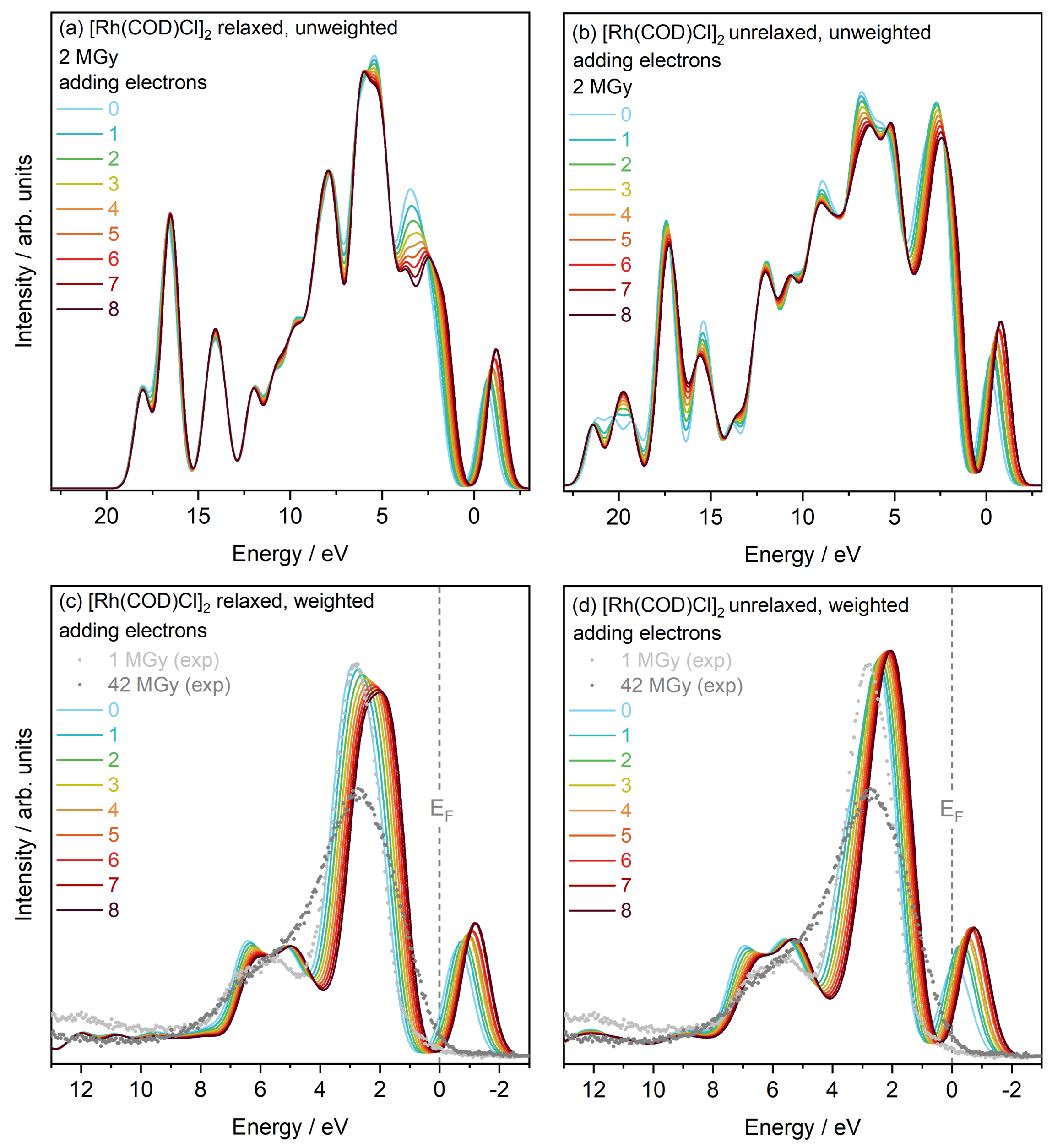}
\caption{\label{fig:Rh_reduction}The calculated density of states for the minimum XRD dose structure of \ce{[Rh(COD)Cl]2} at 2~MGy, mimicking the photoreduction process as 1 to 8 electrons are added to the system. (a) and (b) correspond to the unweighted DOS, for the relaxed and unrelaxed initial structures, respectively. (c) and (d) are the weighted DOS of the relaxed and unrelaxed Rh complex, plotted alongside the minimum and maximum XPS dose VB spectra for comparison.}  
\end{figure*}

\begin{figure*}[!h]
\centering
\includegraphics[width=0.95\textwidth]{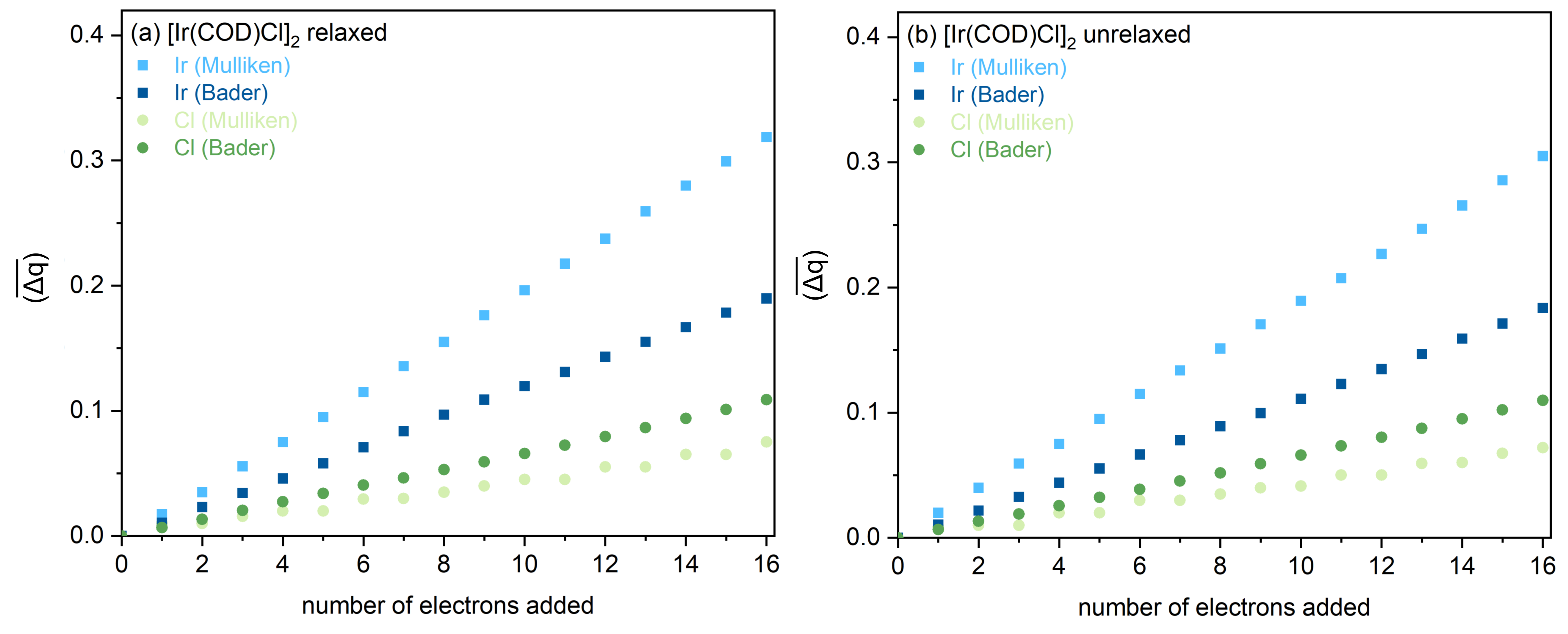}
\caption{\label{fig:Ir_mul_bad}The change in Bader and Mulliken charge of \ce{[Ir(COD)Cl]2}, averaged across all Ir and Cl atoms in the unit cell when electrons are added, relative to the neutral (0 added electrons) charges for the (a) relaxed and (b) unrelaxed, minimum XRD dose structures. }
\end{figure*}

\captionsetup[table]{labelfont=bf,textfont=normalfont}

\begin{figure*}[!h]
\centering
\includegraphics[width=\textwidth]{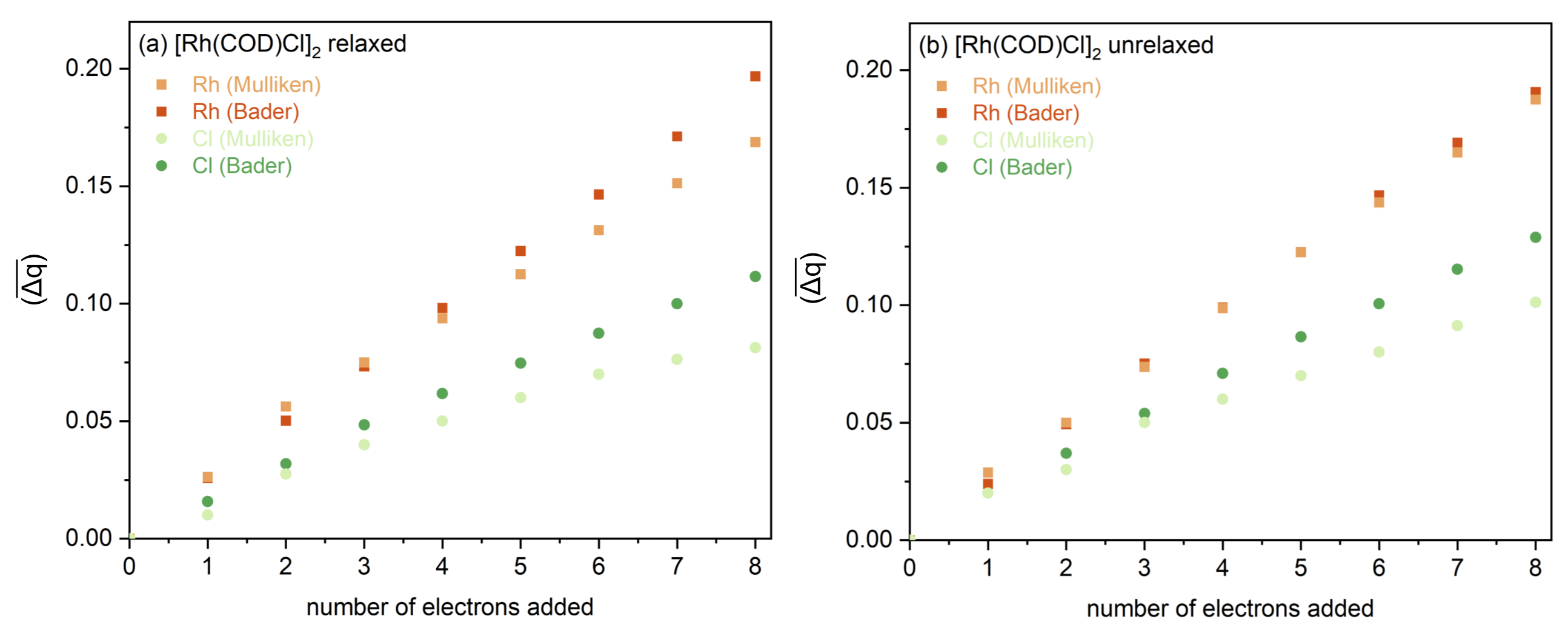}
\caption{\label{fig:Rh_mul_bad}The change in Bader and Mulliken charge of \ce{[Rh(COD)Cl]2}, averaged across all Rh and Cl atoms in the unit cell when electrons are added, relative to the neutral (0 added electrons) charges for the (a) relaxed and (b) unrelaxed, minimum XRD dose structures.}
\end{figure*}

